\documentclass[final]{agujournal2019}

\usepackage{url}
\usepackage{graphicx}
\usepackage{mathtools}
\usepackage{amssymb}
\usepackage{amsmath}
\usepackage{mathrsfs}
\usepackage{xspace}
\usepackage{color}

\journalname{Journal of Geophysical Research: Solid Earth}

\newcommand{\tensor}[1]{\boldsymbol{#1}}
\newcommand{\Grad}{\nabla}
\newcommand{\Div}{\Grad\cdot}

\newcommand{\pd}[2]{\frac{\partial{#1}}{\partial{#2}}}

\newcommand{\vect}[1]{\mathbf{#1}}

\begin{document}

\title{Effects of microstructural heterogeneity on the macroscopic spectrum of elastically accommodated grain-boundary sliding}

\authors{Zhengxuan Li\affil{1} and John F. Rudge\affil{1}}

\affiliation{1}{Bullard Laboratories, Department of Earth Sciences, University of Cambridge, Cambridge, UK}

\correspondingauthor{Zhengxuan Li}{zl471@cam.ac.uk}

\begin{keypoints}
\item Finite-element simulations show irregular grain geometry and grain-size variation modify EAGBS strength but cause little spectrum broadening
\item Broad grain-boundary viscosity distributions suppress the EAGBS peak into a broad attenuation and dispersion background
\item Distributed-viscosity EAGBS can produce asthenosphere-like velocity reduction and attenuation 
\end{keypoints}

\begin{abstract}
Elastically accommodated grain-boundary sliding (EAGBS) is a plausible source of upper-mantle seismic attenuation and dispersion, yet classical theory predicts a localized Debye-like peak that is absent or only weakly expressed in dry olivine experiments. Here we test whether microstructural heterogeneity can explain this discrepancy using 2-D finite-element simulations on periodic Voronoi tessellations. We find that irregular grain geometry changes the baseline EAGBS response relative to the regular hexagonal benchmark, but increasing grain-size variance alone produces only modest changes in modulus and peak height, with little spectral broadening. In contrast, a broad distribution of grain-boundary viscosities progressively suppresses and broadens the Debye-like loss peak into a weak background spanning a wide frequency interval. This broadening arises from the superposition of many localized relaxation processes with distinct characteristic timescales and motivates a reduced-order 0-D description of the aggregate response. These results suggest that the absence of a pronounced EAGBS peak in dry olivine does not necessarily imply the absence of EAGBS mechanism itself. If grain boundaries sample a sufficiently broad viscosity distribution, the macroscopic EAGBS contribution may appear experimentally only as part of a broad attenuation background, while still remaining relevant for upper-mantle seismic attenuation and velocity dispersion.
\end{abstract}

\section*{Plain Language Summary}
Seismic waves lose energy and change speed as they travel through the Earth's upper mantle, but the microscopic processes responsible for these effects are still debated. One possible process is elastically accommodated grain-boundary sliding, where neighboring mineral grains slide slightly past each other while the grains themselves remain elastic. Classical theory predicts that this mechanism should produce a clear peak in seismic attenuation. However, such a peak is weak or absent in experiments on dry olivine, the main mineral of the upper mantle.

Here we use numerical simulations to test whether this missing peak could be hidden by microscopic variability. We find that differences in grain size alone do not strongly broaden the attenuation signal. In contrast, if different grain boundaries have different viscosities, different parts of the rock relax at different rates. Their combined effect spreads the attenuation over a wide frequency range, producing a weak background rather than a sharp peak. This result suggests that grain-boundary sliding may remain important for upper-mantle seismic attenuation and velocity dispersion, even when no distinct attenuation peak is observed.

\section{Introduction}
Seismic data is heavily relied upon to constrain the thermal distribution and physical structure of the Earth's mantle. In hot mantle rocks, seismic propagation is influenced by both elastic and anelastic properties \cite{jacksonGrainsizesensitiveViscoelasticRelaxation2010}.  Seismological models frequently employ a weakly frequency-dependent power-law relationship to describe attenuation. While this approximation holds over typical seismic frequency bands ($10^{-3} - 10^{-1}$ Hz), recent works suggest that the model fails when extrapolated to infinite frequencies \cite{karatoCausalityItsImplications2025}. Theoretical studies have attributed this power-law behaviour to transient diffusion creep, also termed diffusionally-assisted or accommodated grain-boundary sliding (DAGBS). However the predicted seismic attenuation is approximately one order of magnitude lower than observed in the asthenosphere \cite{rudgeViscoelasticRheologyTransient2026}. Therefore, it is critical to identify the micro-physical mechanism responsible for seismic wave speed reduction and attenuation. Elastically accommodated grain boundary sliding (EAGBS) is widely considered to be one of the primary candidate mechanisms driving seismic attenuation and dispersion in the upper mantle \cite{karatoCausalityItsImplications2025}.

Previous theoretical and numerical studies of EAGBS used idealized grain geometries, including bicrystals \cite{leeAnelasticityGrainBoundary2010} and regular hexagonal tessellations \cite{ghahremaniEffectGrainBoundary1980}. Real rocks, however, contain irregular grain shapes and variations in both grain size and grain-boundary properties. Here we use irregular periodic multigrain geometries to quantify network-scale relaxation strengths and spectral widths for comparison with laboratory observations.

Moreover, there has been a longstanding discrepancy between theoretical predictions and experimental observations in synthetic rocks. EAGBS is expected to produce a Debye peak in the attenuation spectrum predicted by the classic Raj-Ashby theory \cite{rajGrainBoundarySliding1971}. Such peaks are repeatedly observed in experiments on metals \cite{keExperimentalEvidenceViscous1947} and ice \cite{tatibouetSTUDYGRAINBOUNDARIES1987, coleCyclicLoadingSaline1995}. A similar feature was also noticed in partially molten rocks \cite{jacksonShearWaveAttenuation2004}. However, dry olivine synthetic rocks only exhibit a negligibly weak peak buried in background attenuation attributed to transient diffusion creep \cite{quOnsetAnelasticBehavior2024,rudgeViscoelasticRheologyTransient2026}. Wet olivine, on the other hand, exhibits a much more prominent peak-like feature \cite{liuEffectWaterSeismic2023}, whose interpretation remains debated between bulk hydrogen-defect relaxation and grain-boundary-related mechanisms \cite{liuEffectWaterSeismic2023,karatoCausalityItsImplications2025}.

Here we test whether microstructural heterogeneity can reconcile the classical EAGBS prediction with the weak peak observed in dry olivine. Namely, we examine the effects of grain-size heterogeneity and grain-boundary viscosity heterogeneity separately on the macroscopic attenuation spectrum, using 2-D finite-element simulations on periodic multi-grain Voronoi tessellations. The physical motivation is that the large crystallographic diversity of olivine grain boundaries may generate a correspondingly broad distribution of grain-boundary transport properties and effective viscosities \cite{marquardtStructureCompositionOlivine2018,wagnerAnisotropySelfdiffusionForsterite2016}. We then use the numerical results to motivate a reduced-order 0-D rheological model in which the macroscopic response is represented as a weighted superposition of localized relaxation processes. Our central result is that grain-size heterogeneity alone has little effect on spectral broadening, whereas a broad distribution of grain-boundary viscosities can transform a localized EAGBS peak into a weak, distributed attenuation background.

The remainder of this manuscript is organized as follows. Section 2 outlines the model setup and numerical framework. Section 3 examines the effects of geometric heterogeneity, and Section 4 considers the effects of grain-boundary viscosity heterogeneity. Section 5 discusses the mechanistic interpretation of the results, develops a reduced-order 0-D rheological model, and explores the implications for laboratory observations and upper-mantle seismology. Section 6 lists the main conclusions from this study. Mathematical and numerical details, including the governing equations, non-dimensionalization, weak formulation, boundary conditions and post-processing, are provided in the appendices.

\section{Model Setup}

\begin{figure}[htbp]
    \centering
    \includegraphics[width=1.0\linewidth]{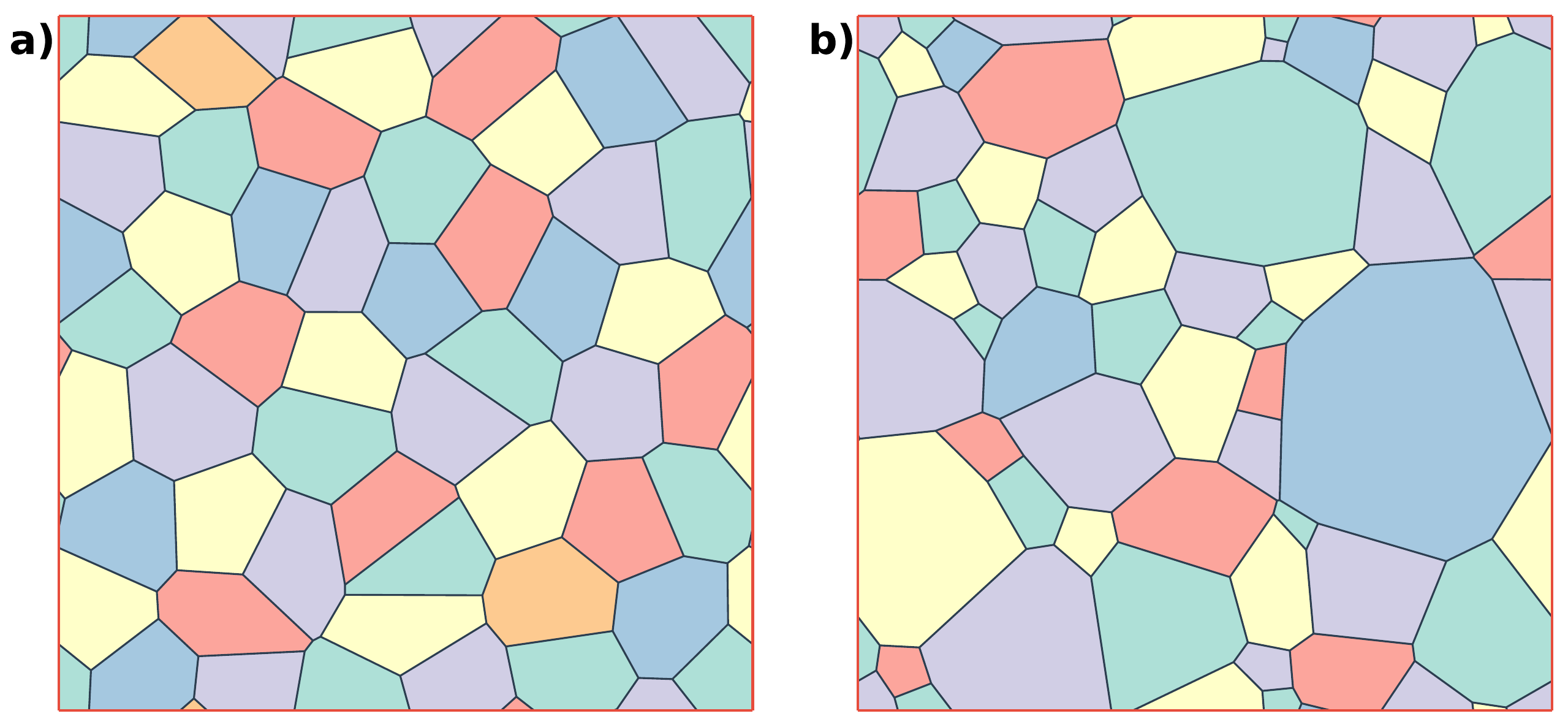}
    \caption{\textbf{Representative polycrystalline RVEs with different grain-size distributions.}
    The two panels show examples of periodic Voronoi tessellations used in the finite-element simulations, with grain sizes prescribed by a log-normal distribution of effective diameter with standard deviation $\sigma_a$. The domain contains 50 grains, with distinct colors indicating different grains (periodic images of the same grain share the same color). Dark lines denote internal grain boundaries, while red lines mark the domain boundaries where periodicity is enforced. Panel (a) shows a nearly uniform grain-size distribution with $\sigma_a$ = 0.05, while panel (b) shows a broader distribution with $\sigma_a$ = 0.5. These RVEs are used to test how grain-size heterogeneity affects the macroscopic EAGBS response. }
    \label{fig:exampleRVE}
\end{figure}

The computational domain is a 2-D periodic polycrystalline aggregate composed of polygonal grains generated from Voronoi tessellations with prescribed grain-size and shape statistics. An example representative volume element (RVE) is illustrated in Figure~\ref{fig:exampleRVE}.

Mechanically, the grain interiors are modeled as isotropic linear elastic solids, while grain boundaries are represented as infinitesimally thin viscous interfaces obeying a Newtonian sliding law. Under a macroscopic, time-harmonic shear deformation, strain energy is stored elastically within the grain interiors and dissipated viscously via intergranular sliding. The system contains an intrinsic time scale, the sliding time, given by:
\begin{equation}
    \tau_e = \frac{\eta_0 a}{\delta \mu}, \label{eq:sliding_time}
\end{equation}
where $\eta_0$ is the reference grain boundary viscosity, $a$ is the characteristic grain size, $\delta$ is the thickness of the viscous layer and $\mu$ is the shear modulus of the grain interior. 

Only tangential viscous sliding is permitted: normal displacement is continuous across every boundary, and no vacancy diffusion or other diffusive matter transport is included. These choices distinguish the present model from the full Raj--Ashby treatment, in which diffusion permits normal boundary motion. The model therefore isolates EAGBS and does not describe the lower-frequency DAGBS regime. The distinction between the two regimes is long established \cite{rajGrainBoundarySliding1971,ghahremaniEffectGrainBoundary1980,jacksonElasticallyAccommodatedGrainboundary2014}. A simple model of grain-boundary viscosity due to \citeA{ashbyBoundaryDefectsAtomistic1972} gives $\tau_e/\tau_M \sim b^2/d^2$, where $\tau_M$ is the Maxwell time, $b\sim10^{-10}$~m is the atomic size and $d$ is the grain size \cite{jacksonGrainsizesensitiveSeismicWave2002,takeiTemperatureGrainSize2014}, placing their characteristic times many orders of magnitude apart. This separation motivates treating EAGBS on its own timescale in the present study.

To quantify this anelastic response, we impose a time-harmonic macroscopic strain and compute the resulting time-harmonic stress. In the Fourier (frequency) domain, the effective macroscopic constitutive relation is given by:
\begin{equation}	
    \hat{\tensor{\sigma}}(\omega) = \mathbb{C}^*(\omega) : \hat{\tensor{\varepsilon}}(\omega),
    \label{eq:constitutive_relationship}
\end{equation}
where hatted quantities denote complex Fourier amplitudes, and $\mathbb{C}^*$ is the effective complex fourth-rank stiffness tensor. For a statistically isotropic 2-D tessellation with no dominant spatial fabric, $\mathbb{C}^*$ is characterized by two independent parameters: the effective bulk modulus $K$ and the effective shear modulus $G$. In the present model, isotropic bulk deformation does not generate grain-boundary sliding. Accordingly, the bulk response remains purely elastic. The effective bulk modulus is therefore purely real, and the only mechanical property requiring numerical evaluation is the complex shear modulus, $G^*(\omega) = G'(\omega) - iG''(\omega)$, where $G'$ and $G''$ denote the storage and loss moduli, respectively. For each simulated microstructure and angular frequency $\omega$, we solve the corresponding boundary-value problem by the finite element method and extract the effective complex modulus from cycle-averaged stored and dissipated energies. The grain geometries are generated using Neper \cite{queyOptimalPolyhedralDescription2018}, the finite-element discretizations and solution were realized using NGSolve and the mesh was generated using Netgen \cite{schoeberlNGSolveNetgenV6226012026}. Repeating this over a range of frequencies yields the macroscopic attenuation spectrum. Mathematical details of the governing equations, weak formulation, periodic boundary conditions and post-processing are provided in \ref{App:equations} to \ref{App:Homogenisation}.

We focus on two sources of microstructural heterogeneity in this study: variation in grain size and variation in grain-boundary (GB) viscosity. Grain-size heterogeneity modifies the geometric length scales over which stresses are transmitted and redistributed throughout the aggregate, thereby affecting patterns of strain localization. GB-viscosity heterogeneity alters the local resistance to intergranular sliding, directly controlling the relaxation rates of different parts of the boundary network. In both cases, the aggregate samples a distribution of effective relaxation times rather than a single characteristic time scale. At the local level, this relaxation time is expected to scale approximately as $\tau_i \propto \eta_{\mathrm{gb},i}/k_i \propto \eta_{\mathrm{gb},i}a_i$, where $\eta_{\mathrm{gb},i}$ is the local boundary viscosity, $k_i$ is an effective local stiffness dictated by the surrounding grain geometry, which is inversely proportional to the local grain size $a_i$. This distribution of $\tau_i$  controls the frequency-dependent dispersion in $G'(\omega)$ and the breadth of the attenuation peak in $G''(\omega)$.

In the following sections, we first vary the grain-size distribution while holding GB viscosity uniform, and then vary the GB-viscosity distribution while holding the grain geometry fixed, so that the distinct effects of geometric and rheological heterogeneity on the shear attenuation spectrum can be assessed separately.

\section{Effect of Geometric Heterogeneity}

To investigate the contributions of geometric heterogeneity to the macroscopic attenuation spectrum, we computed the mechanical response of polycrystalline aggregates with varying grain-size distributions while holding the intrinsic grain-boundary properties constant. The microstructures were generated as 2-D periodic Voronoi tessellations in which the effective grain diameter, $a$, follows a log-normal distribution, $\ln a \sim \mathcal{N}(\mu_a,\sigma_a^2)$, where $\mu_a$ is the logarithmic mean and $\sigma_a$ controls the width of the distribution.

To ensure statistical robustness, 100 independent sample geometries were generated for each $\sigma_a$, and the resulting complex shear moduli were ensemble-averaged to obtain the macroscopic $G'(\omega)$ and $G''(\omega)$ spectra. With 100 realizations, the standard error of the mean on the peak attenuation is below 1\% for all $\sigma_a$ levels, confirming that this ensemble size is sufficient to resolve the reported trends. Across all simulations in this suite, the non-dimensionalized grain-boundary viscosity was held fixed at unity, $\eta_{\mathrm{gb}} = 1$. In addition, to isolate the effect of grain-size variance from that of the absolute grain scale, the geometric mean of effective grain diameter was held constant as
$\sigma_a$ was varied. The resulting effective moduli are normalized by the  elastic shear modulus of the grain interiors, $\mu$.

\subsection{Benchmark and the Role of Irregular Grain Shapes}

\begin{figure}[htbp]
    \centering
    \includegraphics[width=\linewidth]{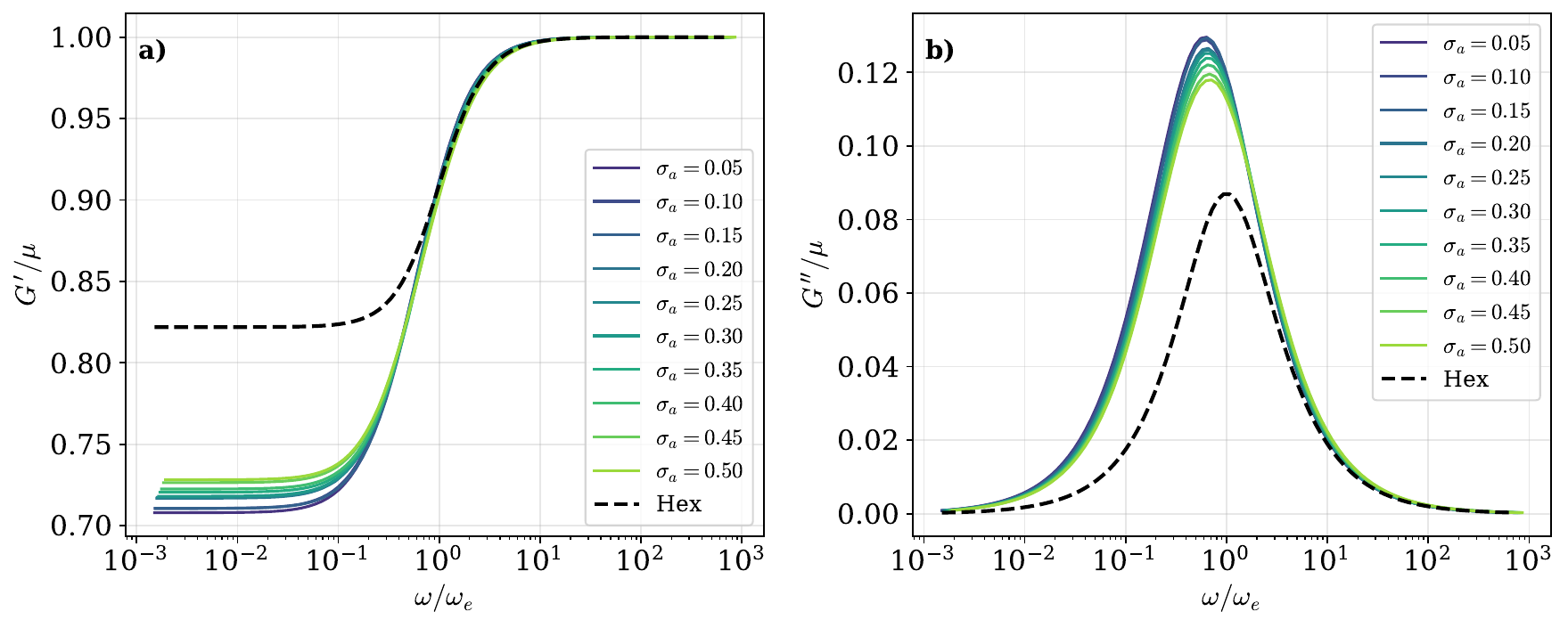}

    \caption{\textbf{Macroscopic effective moduli as functions of normalized angular frequency ($\omega/\omega_e$), illustrating the effects of microstructural topology and grain-size variance.} Here, $\omega_e$ is the frequency at which $G''$ peaks for the regular-hexagon RVE. \textbf{(a)} Normalized storage modulus ($G'/\mu$). \textbf{(b)} Normalized loss modulus ($G''/\mu$). The dashed black line shows the benchmark response of a perfectly regular hexagonal tessellation, reproducing the results of Ghahremani (1980). The solid colored lines show ensemble-averaged responses of Voronoi tessellations with increasing standard deviation $\sigma_a$ of the log-normal grain-size distribution.}
    \label{fig:grain_size_spectra}
\end{figure}

To validate the numerical framework, we first computed the macroscopic response of a perfectly regular hexagonal tessellation (Figure~\ref{fig:grain_size_spectra}, dashed black lines). Here and in all later simulations, a Poisson's ratio of the grain interiors $\nu = 0.35$ was adopted as in \citeA{ghahremaniEffectGrainBoundary1980}. The resulting spectra reproduce the analytical and numerical results of \citeA{ghahremaniEffectGrainBoundary1980}. In this idealized geometry, all grain boundaries have identical lengths and all triple junctions meet at symmetric $120^{\circ}$ angles, strongly constraining intergranular sliding. As a result, the aggregate exhibits a relatively high relaxed storage modulus ($G'/\mu \approx 0.82$ as $\omega \rightarrow 0$) and a single narrow Debye-like attenuation peak with $G''_{\max}/\mu \approx 0.09$. It is common to define the relaxation strength $\Delta$ as:
\begin{equation}
    \Delta = 1-\frac{G'}{\mu} \quad \text{for} \quad \omega \rightarrow 0.
\end{equation} 
In this case the calculated relaxation strength is $\Delta = 0.18$.

We then replaced this idealized benchmark with irregular polygonal grains generated from a Voronoi tessellation with a narrow grain-size distribution ($\sigma_a = 0.05$). As shown in Figure~\ref{fig:grain_size_spectra}a, the introduction of randomized boundary lengths and irregular junction angles allows greater intergranular sliding, leading to a larger relaxed compliance. Consequently, the relaxation strength rises sharply to $\Delta \approx 0.28$. 

In addition to increasing the total relaxation strength, the transition from regular hexagons to irregular polygonal grains systematically shifts the attenuation peak to lower frequency by a factor of approximately $1.4$. This offset remains largely unchanged as $\sigma_a$ is increased in the subsequent simulations, indicating that it is not controlled by the width of the grain-size distribution itself. It therefore appears to arise mainly from grain-shape and network-topology irregularity relative to the regular hexagonal benchmark.

\subsection{Influence of Grain-Size Variance}

Having established the strong effect of irregular grain geometry, we next investigate the role of grain-size variance within the Voronoi aggregates by increasing the standard deviation of the log-normal grain-size distribution from $\sigma_a = 0.05$ to $\sigma_a = 0.50$.

As shown in Figure~\ref{fig:grain_size_spectra}, broadening the grain-size distribution produces only a modest change in the macroscopic spectra. In Figure~\ref{fig:grain_size_spectra}a, increasing $\sigma_a$ produces a slight increase in the relaxed modulus. Correspondingly, Figure~\ref{fig:grain_size_spectra}b shows a minor but systematic reduction in the peak attenuation height. However, the loss spectrum remains narrow and retains a Debye-like form across the full range of $\sigma_a$ examined.

Increasing grain-size variance does not produce appreciable spectral broadening, nor does it significantly alter the frequency shift introduced by the transition from regular to irregular grain geometry. Thus, although grain-size heterogeneity slightly modifies the relaxation strength and peak position, it does not broaden or suppress the attenuation peak to account for the weak or poorly resolved peak observed in dry olivine experiments. Within the present model framework, geometric heterogeneity alone is therefore insufficient to explain the experimental discrepancy.

\section{Effect of Grain-Boundary Viscosity Distribution}

Having demonstrated that geometric heterogeneity alone is insufficient to explain the weak peak in dry olivine experiments, we now investigate the role of grain-boundary (GB) viscosities. In orthorhombic minerals like olivine, the grain-boundary character space is highly diverse, leading to transport properties that can vary by orders of magnitude depending on boundary character and crystallographic misorientation \cite{marquardtStructureCompositionOlivine2018,wagnerAnisotropySelfdiffusionForsterite2016}. Here we represent this structural and chemical diversity by a log-normal distribution of GB viscosities ($\eta_{\mathrm{gb}}$) and test whether such a distribution broadens the macroscopic EAGBS response.

\subsection{Macroscopic Spectral Broadening}

To test this hypothesis, we assigned each internal grain boundary an independently sampled viscosity factor in our 2-D Voronoi aggregates, drawn from a log-normal distribution centered on the geometric mean viscosity $\bar{\eta}$: $\ln(\eta_{\mathrm{gb}}/\bar{\eta}) \sim \mathcal{N}(0, \sigma_{\eta}^2)$. The value of $\bar{\eta}$ was held fixed so that the characteristic relaxation frequency remained within the numerical frequency window, and the distribution width ($\sigma_{\eta}$) was varied from 0.10 (near-uniform) to 7.00 (highly heterogeneous). Each spectrum shown in this section is an ensemble average over 100 Voronoi geometries generated with a grain-size standard deviation of $\sigma_a=0.05$. Similar peak suppression and broadening were obtained when the calculations were repeated for the other $\sigma_a$ values examined in Section~3, consistent with the weak sensitivity to grain-size variance established there.

\begin{figure}[htbp]
    \centering
    \includegraphics[width=\textwidth]{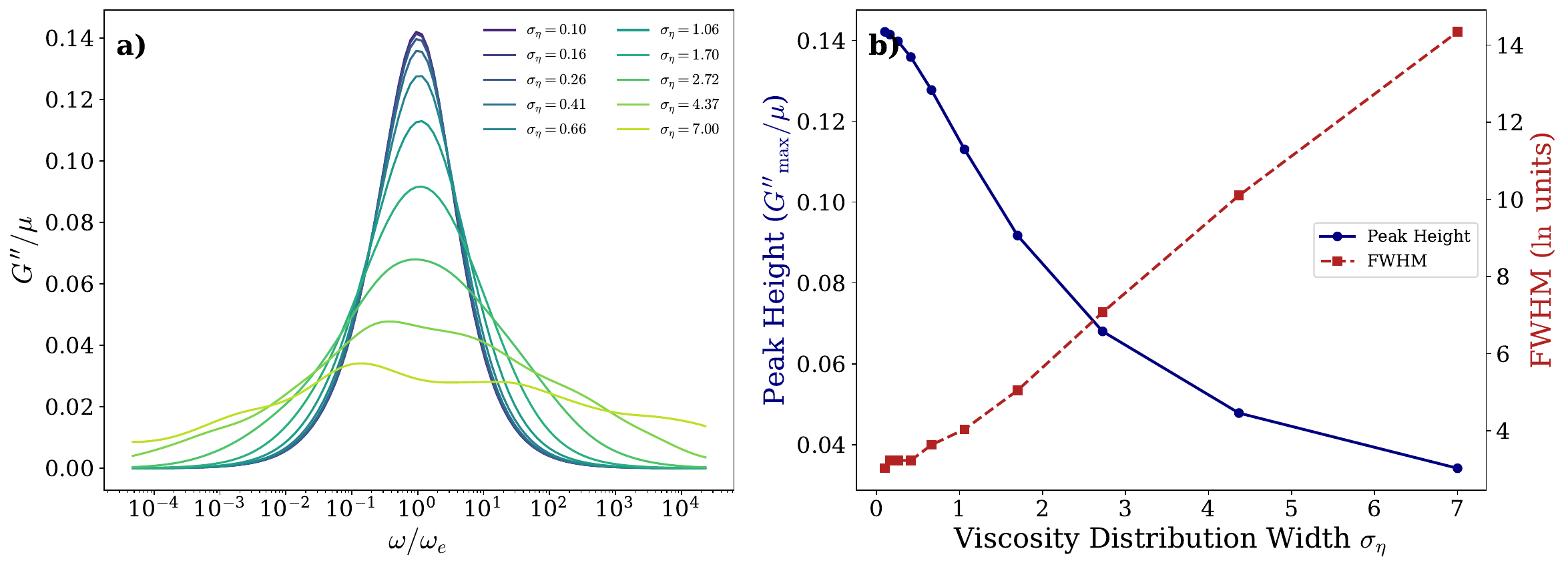}
    \caption{\textbf{Evolution of the macroscopic attenuation spectrum as a function of grain-boundary viscosity heterogeneity.} \textbf{(a)} Ensemble-averaged loss modulus ($G''/\mu$) versus normalized angular frequency ($\omega/\omega_e$) for varying log-normal distribution widths ($\sigma_\eta$). For this figure, $\omega_e$ is the peak frequency of the otherwise identical near-uniform Voronoi reference calculation with $\sigma_\eta=0.01$. \textbf{(b)} Quantitative trends of the spectral reshaping, illustrating the monotonic decrease in peak height (blue circles, left axis) and the corresponding increase in the full-width at half-maximum (FWHM; red squares, right axis) as the distribution widens.}
    \label{fig:spectra_trends}
\end{figure}

The resulting macroscopic attenuation spectra, computed as the ensemble-averaged loss modulus $G''/\mu$, are presented in Figure \ref{fig:spectra_trends}a. As the variance in boundary viscosity increases, the localized Debye-like peak systematically flattens, progressively broadening into a weaker, more distributed background. Figure \ref{fig:spectra_trends}b quantifies this transition: as $\sigma_{\eta}$ increases, the peak height decreases monotonically while the full-width at half-maximum (FWHM) increases approximately linearly.

\subsection{Mechanistic Origin of the Broad Background}

\begin{figure}[htbp]
    \centering
    \includegraphics[width=\textwidth]{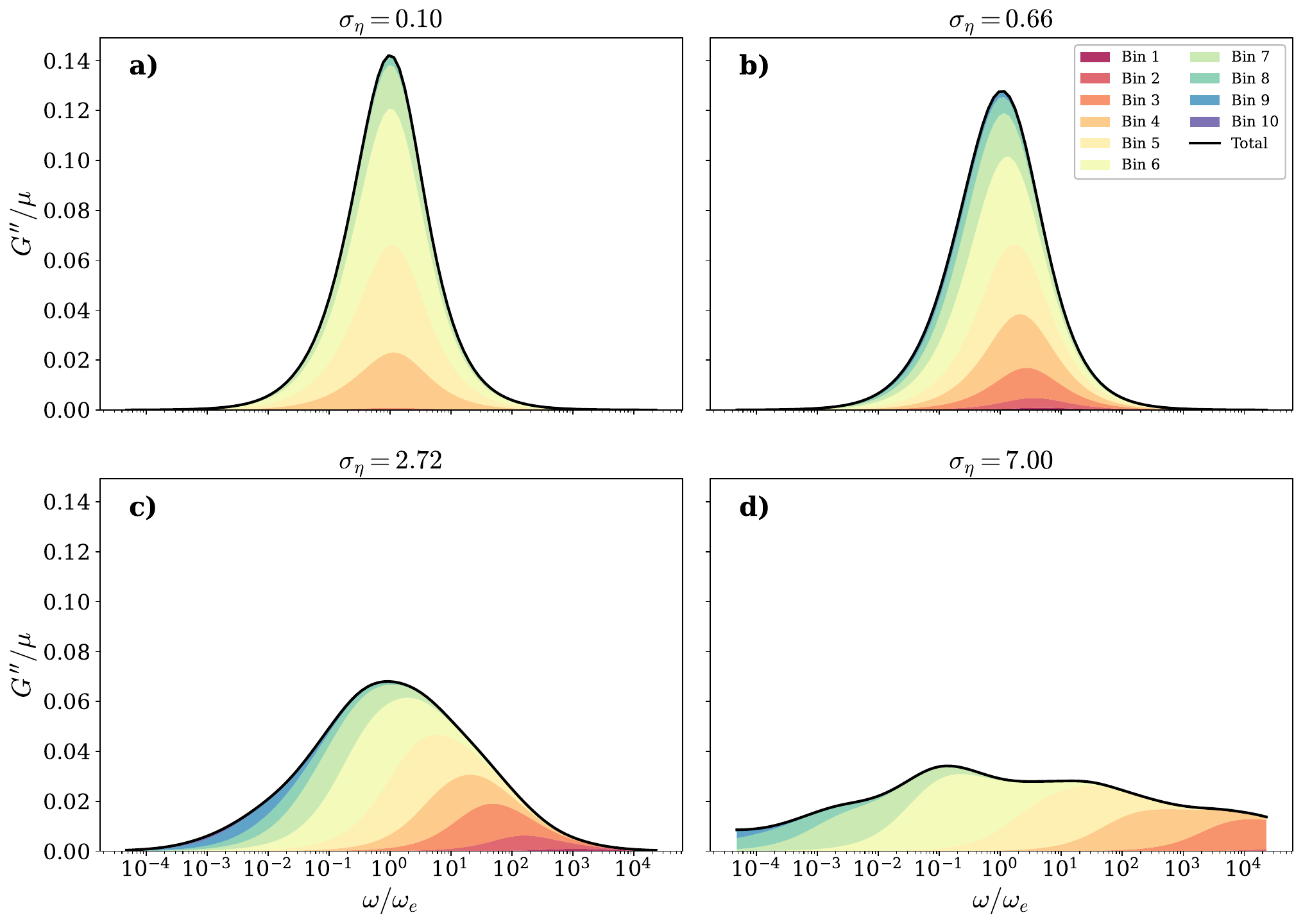}
    \caption{\textbf{Mechanistic decomposition of macroscopic attenuation into discrete viscosity classes.} Representative panels demonstrate the transition from a near-uniform viscosity distribution ($\sigma_\eta = 0.1$) to a highly heterogeneous one ($\sigma_\eta = 7.0$). In each panel, the total macroscopic attenuation (solid black line) is decomposed into the additive contributions of 10 log-spaced grain-boundary viscosity bins (colored stacked areas). At large $\sigma_\eta$, the contributions from different viscosity bins separate along the frequency axis, and the total spectrum broadens accordingly. Note that the curve in panel (d) appears less smooth because the finite number of grain boundaries only sparsely samples the viscosity distribution when $\sigma_\eta$ is very large.}
    \label{fig:decomposition}
\end{figure}

To examine the origin of this spectral broadening, we grouped grain boundaries into ten log-spaced bins according to their viscosities and decomposed the total macroscopic dissipation into additive contributions from each bin (Figure~\ref{fig:decomposition}). Details of the decomposition procedure are given in \ref{App:Homogenisation}.

At near-uniform distributions ($\sigma_{\eta} = 0.10$), all viscosity bins relax nearly simultaneously, superimposing to form a single, narrow Debye-like peak. However, as $\sigma_{\eta}$ increases, the individual bin responses separate along the frequency axis. Each viscosity class retains a localized Debye-like response, and the broad macroscopic spectrum arises from their superposition over distinct characteristic frequencies.

\subsection{Additivity of Integrated Grain-Boundary Dissipation}

Although the full 2-D simulations resolve a mechanically coupled network of grains, boundaries, and triple junctions, it is useful to determine whether the integrated dissipation is controlled primarily by network connectivity or by the total abundance of each viscosity class.

\begin{figure}[htbp]
    \centering
    \includegraphics[width=\textwidth]{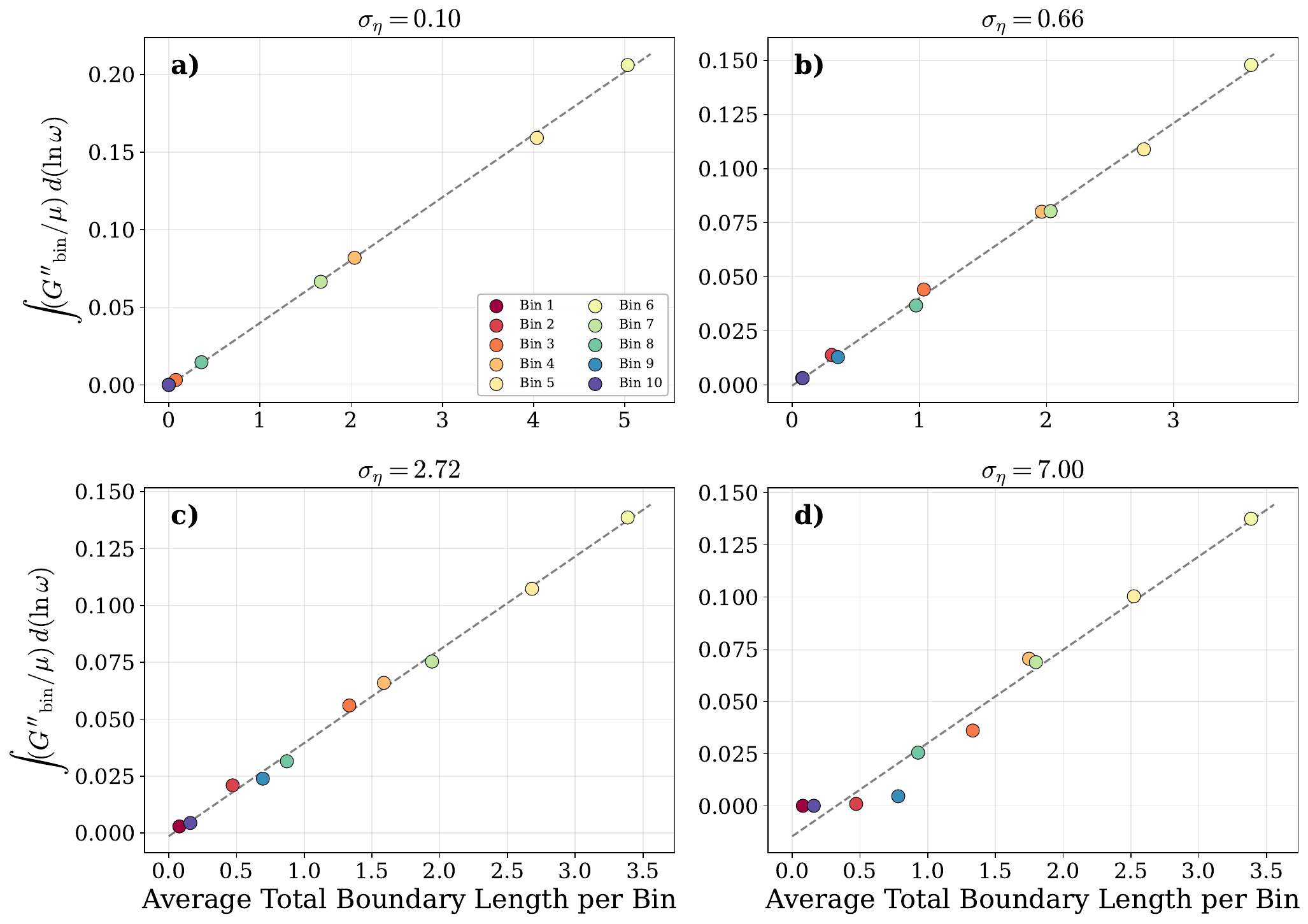}
    \caption{\textbf{Additivity of macroscopic dissipation.} The integrated dissipation (the area under the $G''/\mu$ curve) for each independent viscosity bin is plotted against its normalized fractional boundary length within the aggregate. To leading order, the total energy dissipated by a specific viscosity class is directly proportional to its total length in the microstructural network.}
    \label{fig:scaling_law}
\end{figure}

To assess this, we integrated each bin-resolved loss spectrum over frequency and compared the resulting dissipation with the total boundary length of the corresponding viscosity class within the aggregate (Figure~\ref{fig:scaling_law}). For most distribution widths, the two quantities exhibit an approximately linear relationship.

This scaling implies that, to leading order, the macroscopic dissipation contributed by a given viscosity class follows its total fractional representation in the grain-boundary network. In other words, although the instantaneous stress and sliding fields remain fully coupled through the elastic grain framework, the integrated dissipation can be partitioned approximately additively across viscosity classes. The dominant control on the total energy dissipated by a given class is therefore its total boundary length, rather than higher-order details of boundary connectivity.

Small deviations from linearity become more apparent only at the largest distribution widths ($\sigma_\eta \gtrsim 5$). In these cases, the characteristic relaxation frequencies of the most viscous and least viscous bins are shifted toward or beyond the edges of the numerical frequency window, so their integrated areas are slightly underestimated. These departures are therefore attributable to finite frequency-window truncation rather than a signature of non-linearity.

This result has two important implications. First, it shows that broad aggregate-scale attenuation can be interpreted as the cumulative contribution of many localized relaxations associated with different grain-boundary viscosities. Second, it provides the numerical motivation for a reduced-order rheological model in which the macroscopic EAGBS spectrum is approximated by a weighted superposition of localized Debye-like responses.

\section{Discussion}

\subsection{A Reduced-Order (0-D) Rheological Model for EAGBS with Distributed Viscosity} \label{sec:0-D}

Taken together, the results of Sections~3 and~4 confirm the qualitative contrast identified by \citeA{leeAnelasticityGrainBoundary2010}. In their prescribed saw-tooth bi-crystal, a fourfold grain-size variation weakened the loss peak with little change in shape, whereas an order-of-magnitude contrast in boundary viscosity both weakened and broadened it. The present calculations show that this distinction persists across a coupled periodic multi-grain network with continuous log-normal distributions and ensembles of irregular geometries. In addition, we quantify the geometry-dependent relaxation strength, the progressive change in spectral width, and the approximately additive integrated dissipation. This additivity of macroscopic dissipation suggests that, to leading order, the macroscopic EAGBS response can be approximated as a sum of contributions from different viscosity classes weighted by the boundary length.

This additivity motivates a reduced-order (0-D) representation of the EAGBS spectrum. In this approximation, the macroscopic response is represented as the superposition of many Debye relaxation elements (standard linear solids) connected in parallel, each associated with a grain-boundary viscosity $\eta_{\mathrm{gb}}$. In continuous form, the loss modulus is
\begin{equation}
    G''_{\mathrm{EAGBS}}(\ln{\omega})
    \approx
    \int g''_{\mathrm{single}}(\ln{\omega};\eta_{\mathrm{gb}})
    \, p(\eta_{\mathrm{gb}})
    \,\mathrm{d}(\ln \eta_{\mathrm{gb}}),
    \label{eq:EAGBS_0D_loss}
\end{equation}
where $g''_{\mathrm{single}}(\omega;\eta_{\mathrm{gb}})$ is the localized Debye-like loss response of an individual viscosity class and $p(\eta_{\mathrm{gb}})$ is the corresponding length-weighted probability density function. In practice, the reduced-order model is implemented at the level of the complex modulus, so that both the storage and loss components, $G'(\omega)$ and $G''(\omega)$, are obtained from the real and imaginary parts of the same distributed standard-linear-solid representation. The explicit form used for this calculation is given in Appendix \ref{App: E3}.

Figure~\ref{fig:0D_SLS_model} shows that this simple model reproduces the main qualitative trend seen in the FEM calculations: increasing the width of the viscosity distribution suppresses the peak height and broadens the spectrum into a weak background. 

The 0-D formulation provides both a compact physical interpretation of the numerical results and a computationally efficient way to explore how an assumed grain-boundary viscosity distribution maps into an attenuation spectrum. 

The 0-D model does not eliminate the role of geometry. As shown in Section~3, irregular grain topology changes both the relaxation strength and the characteristic frequency relative to the regular hexagonal benchmark. In the narrow-distribution limit, these geometric effects provide the calibration for the amplitude and frequency scale of the reduced-order model. The present formulation should therefore be understood as separating the problem into two parts: geometric calibration of the relaxation response, and statistical averaging over the grain-boundary viscosity distribution.

\begin{figure}[htbp]
    \centering
    \includegraphics[width=\textwidth]{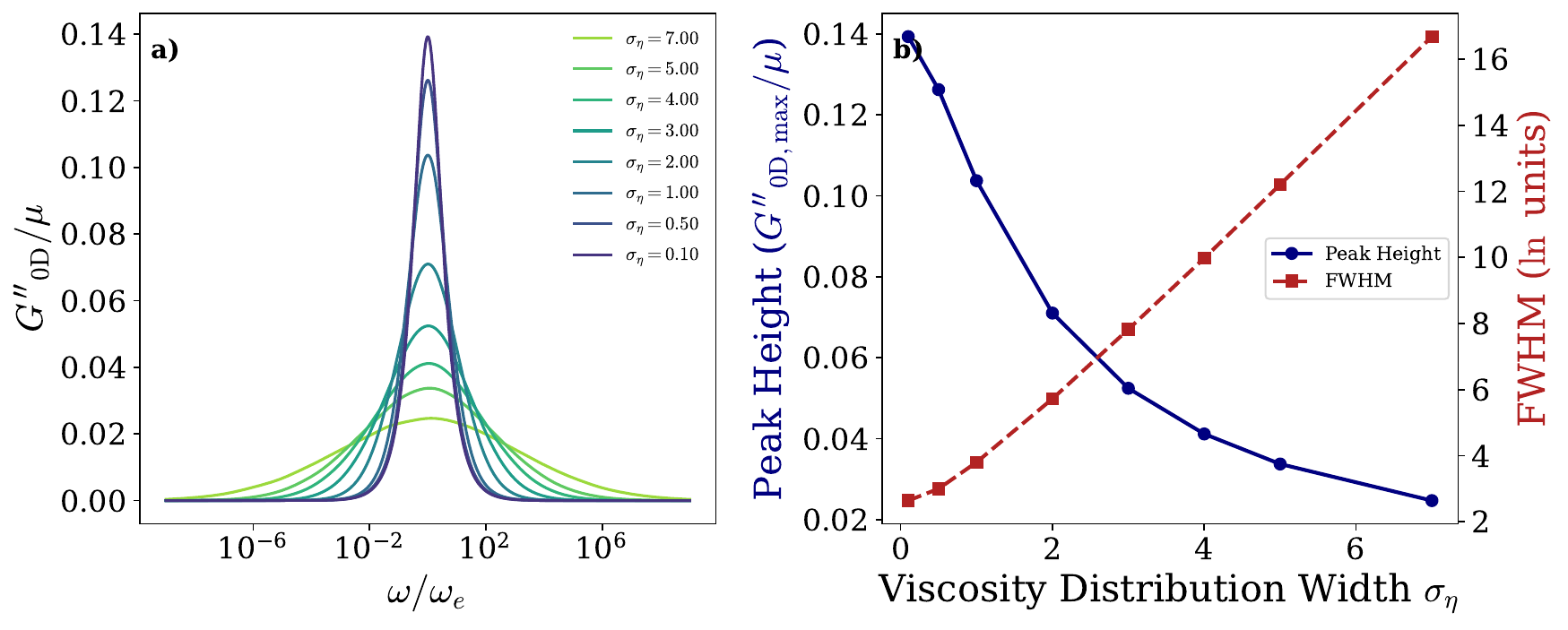}
    \caption{\textbf{0-D SLS model for EAGBS with distributed grain-boundary viscosity.}
    \textbf{(a)} Calculated loss modulus from a reduced-order model, in which the macroscopic response is constructed as the parallel superposition of many independent Debye-like relaxation elements whose viscosities follow a log-normal distribution of width $\sigma_\eta$. The characteristic frequency $\omega_e$ is defined from the peak of the narrowest distribution. 
    \textbf{(b)} Quantitative trends extracted from the 0-D model. Increasing $\sigma_\eta$ produces a monotonic decrease in peak height together with an increase in full-width at half-maximum (FWHM).}
    \label{fig:0D_SLS_model}
\end{figure}

\begin{figure}[htbp]
    \centering
    \includegraphics[width=\textwidth]{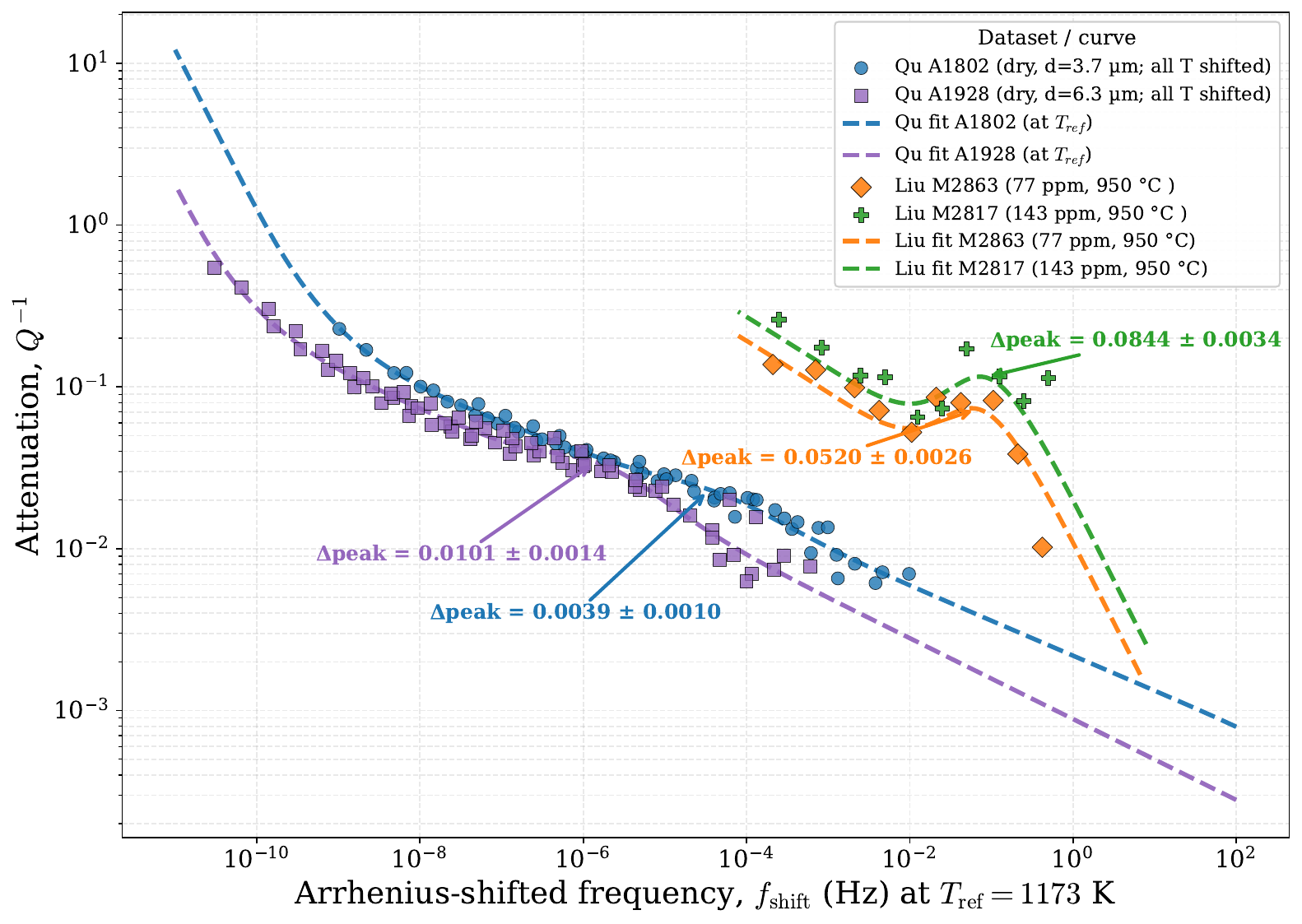}
    \caption{\textbf{Comparison of attenuation spectra for dry and wet olivine on a common shifted-frequency axis.}
    Dry olivine data from \citeA{quOnsetAnelasticBehavior2024} are plotted after collapsing measurements acquired at multiple temperatures onto a common reference temperature $T_{\rm ref}=1173$~K using the sample-specific Arrhenius shift. Wet olivine data from \citeA{liuEffectWaterSeismic2023} are shown together with their fitted curves after shifting to the same reference temperature. Dry olivine is dominated by a broad background with only a subtle peak contribution, whereas wet olivine exhibits a distinctly peak-like feature whose amplitude increases with water content.
    The plotted curves reproduce the published fits of the original authors and are not fits of the distributed-viscosity EAGBS model developed here. In those fits, the peak-like contribution is represented empirically by a Gaussian distribution in $\ln\tau$ (a log-normal peak component), and the DAGBS contribution by a generalized-Burgers absorption-band background. The annotated $\Delta_{\rm peak}$ values are peak heights, defined as the maximum excess in $Q^{-1}$ above the fitted background. They are related to but are not a direct measure of the relaxation strength. }
    \label{fig:dry_wet_overlay}
\end{figure}
\subsection{Implications for Dry and Wet Olivine Experiments}\label{sec:dry_and_wet}

Figure~\ref{fig:dry_wet_overlay} shows that dry olivine is characterized by a broad attenuation background with only a weak peak-like feature expressed mainly as a change in slope \cite{quOnsetAnelasticBehavior2024}. The results of Sections~4 and~5.1 suggest a possible explanation for this behaviour. If grain boundaries sample a sufficiently broad distribution of viscosities, then the EAGBS contribution is spread over a wide range of relaxation times and may become difficult to distinguish from the background attenuation.

Such a broad distribution is physically plausible in dry olivine. Grain-boundary character in olivine spans a large five-dimensional crystallographic space \cite{marquardtStructureCompositionOlivine2018}. Since olivine is orthorhombic with low crystal symmetry, its grain-boundary character space is far more diverse than that of face-centered cubic crystals such as aluminum.  SEM observations further suggest that olivine grain boundaries do not cluster strongly around a small set of preferred misorientation axes \cite{faulGrainMisorientationsPartially1999}. Molecular-dynamics calculations indicate that grain-boundary transport properties vary by more than three orders of magnitude across this space \cite{wagnerAnisotropySelfdiffusionForsterite2016}. Additional variability in boundary thickness, local defect content, and interfacial structure would broaden the effective viscosity distribution still further. Dry olivine is therefore expected to sample a wide range of grain-boundary viscosities.

For scale, a log-normal distribution with $\sigma_\eta=4$ has a one-standard-deviation interval of $\exp(\pm4)$ about the median, spanning a range of approximately $3\times10^3$, or about 3.5 orders of magnitude, in viscosity. This is comparable to the order-of-magnitude variability suggested by molecular-dynamics calculations, although the precise width of the effective grain-boundary viscosity distribution in dry olivine remains poorly constrained.

An auxiliary comparison with paired modulus and attenuation measurements from dry olivine is given in \ref{app:dry_joint_fit}. It shows that a broad EAGBS contribution can be combined with a DAGBS background without producing a separately resolved EAGBS peak in the measured window. 

The same framework may also explain why pronounced EAGBS peaks are more readily observed in simpler polycrystalline systems. In face-centered cubic metals such as aluminum, systematic surveys of all 388 distinct grain-boundary types find that the spread in mobility is narrower than in olivine \cite{olmstedSurveyComputedGrain2009}. The sliding time is then concentrated around a more localized characteristic timescale, allowing the EAGBS peak to survive macroscopic averaging. This is consistent with classic observations in metals \cite{keExperimentalEvidenceViscous1947}. The contrast with dry olivine is therefore consistent with the broader conclusion of this study: whether EAGBS appears experimentally as a distinct peak or only as a weak background contribution depends strongly on the width and position of the underlying grain-boundary viscosity distribution.

The wet olivine measurements of \citeA{liuEffectWaterSeismic2023} present the opposite case. In Figure~\ref{fig:dry_wet_overlay}, their spectra show a localized peak-like feature that is both narrower and larger in amplitude than anything resolved in the dry data, and whose amplitude increases with water content. Taken at face value, this is the contrast that the present framework would predict if hydration acted to narrow the effective distribution of grain-boundary viscosities. However, the physical mechanism through which this peak emerges remains disputed.

\citeA{liuEffectWaterSeismic2023} did not attribute the peak to grain-boundary sliding. They interpreted it as a bulk relaxation associated with the diffusion of hydrogen-bearing defects, principally because the fitted peak position showed little clear dependence on grain size, whereas the characteristic sliding time in EAGBS is expected to scale linearly with grain size (equation~\ref{eq:sliding_time}). \citeA{karatoCausalityItsImplications2025} contested this inference, arguing that the apparent grain-size independence is not robust and that, after excluding one outlying sample, the fitted peak positions vary with grain size linearly, which is more consistent with a grain-boundary process. 

If the wet peak does reflect grain-boundary sliding, then the present framework requires specifically that hydration narrows the distribution of $\eta_{\mathrm{gb}}$, rather than simply lowering its mean: a uniform reduction in boundary viscosity would translate the spectrum along the frequency axis without sharpening it. There is some independent reason to expect such a narrowing. Atomistic calculations on forsterite show that hydrogen-bearing defects are energetically favoured at grain boundaries and that hydrated boundaries may develop lower-density interfacial regions \cite{deleeuwProtoncontainingDefectsForsterite2000}, suggesting that hydration modifies the boundary network before comparable effects are expressed uniformly throughout the bulk. If the modified boundaries converge on a more weakly structured and homogeneous interfacial state, the broad dry-olivine distribution inferred above would narrow and a Debye-like peak would re-emerge from the background. A qualitatively similar sharpening is observed in melt-bearing olivine aggregates, where a more distinct peak-like relaxation feature is resolved \cite{faulShearWaveAttenuation2004,jacksonShearWaveAttenuation2004}.

This interpretation is also challenged. First, \citeA{quOnsetAnelasticBehavior2024} noted that the centre period of the wet peak shows little temperature dependence. This is difficult to reconcile with any thermally activated relaxation, and therefore counts against the grain-boundary reading as much as against the bulk-diffusion one. Second, \citeA{clineiiRedoxinfluencedSeismicProperties2018} found shear-wave speeds and attenuation to be insensitive to water content across the water-undersaturated range they examined, but that redox conditions exerting the stronger influence, and reported no support for water-enhanced grain-boundary sliding. \citeA{liuEffectWaterSeismic2023} attributed the discrepancy primarily to differences in pressure, composition, and defect chemistry as they used iron-free samples. However, because their samples were iron-free and redox variability was correspondingly suppressed, their experiments cannot separate the effects of water from those of oxygen fugacity in natural iron-bearing olivine. 

These competing interpretations suggest an experimental discriminator. If the wet peak is controlled primarily by a bulk hydrogen-defect relaxation, its strength should continue to increase with water content. If instead it reflects grain-boundary sliding made more visible by preferential boundary hydration, its amplitude should eventually saturate once most boundaries have been modified, since further water uptake would then act on an already-saturated boundary network. To reconcile this with \citeA{clineiiRedoxinfluencedSeismicProperties2018} and \citeA{quOnsetAnelasticBehavior2024}, the experiment would require measurements that independently control water content and oxygen fugacity while also resolving the temperature and grain-size dependence of the peak position.

\subsection{Implications for Seismic Attenuation and Velocity Dispersion}

Upper-mantle viscoelasticity is expressed through both seismic attenuation and velocity dispersion. 
For weak attenuation, the inverse quality factor is approximately
\begin{equation}
    Q^{-1}(\omega) \approx \frac{G''(\omega)}{G'(\omega)},
\end{equation}
and the corresponding shear-wave velocity is
\begin{equation}
    V_s(\omega) \approx \sqrt{\frac{G'(\omega)}{\rho}},
\end{equation}
where $\rho$ is density.

Recent work has argued that transient diffusion creep alone may be insufficient to explain the observed attenuation of the asthenosphere at seismic periods. In particular, \citeA{rudgeViscoelasticRheologyTransient2026} showed that extrapolation of transient diffusion creep to seismic frequencies yields attenuation significantly smaller than the $Q^{-1}\sim10^{-2}$ often inferred from seismology. This suggests that an additional dissipative mechanism may be required under upper-mantle conditions.

In terms of dispersion, \citeA{priestleyThermalAnisotropicStructure2024} parameterized surface-wave velocity as a function of temperature and pressure by combining a low-temperature elastic dependence of density and unrelaxed shear modulus with a high-temperature Andrade-type anelastic reduction. They emphasized that this high-temperature parameterization is empirical, although it provides a simple description of the observed broad power-law anelastic behavior. Here we retain the same low-temperature elastic parameterization, but replace the empirical high-temperature Andrade-type reduction with the distributed-viscosity EAGBS model developed in Section~\ref{sec:0-D}, to examine whether grain-boundary sliding can generate a comparably broad anelastic velocity reduction. The reference grain size and temperature are chosen to match the dry-olivine laboratory state of \citeA{quOnsetAnelasticBehavior2024}; the reference sliding time $\tau_{\mathrm{ref}}$, activation energy $E_a$, and log-normal distribution width $\sigma_\eta$ are fitted to the seismic velocity data rather than independently fixed by those experiments. A full mathematical statement of the fitted reduced-order model, including the definitions of $\mu(T,P)$, $\rho(T,P)$, and $V_{s,u}(T,P)$, together with the inversion procedure and best-fitting parameters, is provided in \ref{app:fitting_model} and Table~\ref{tab:fitting_parameters}.

\begin{figure}[htbp]
    \centering
    \includegraphics[width=\textwidth]{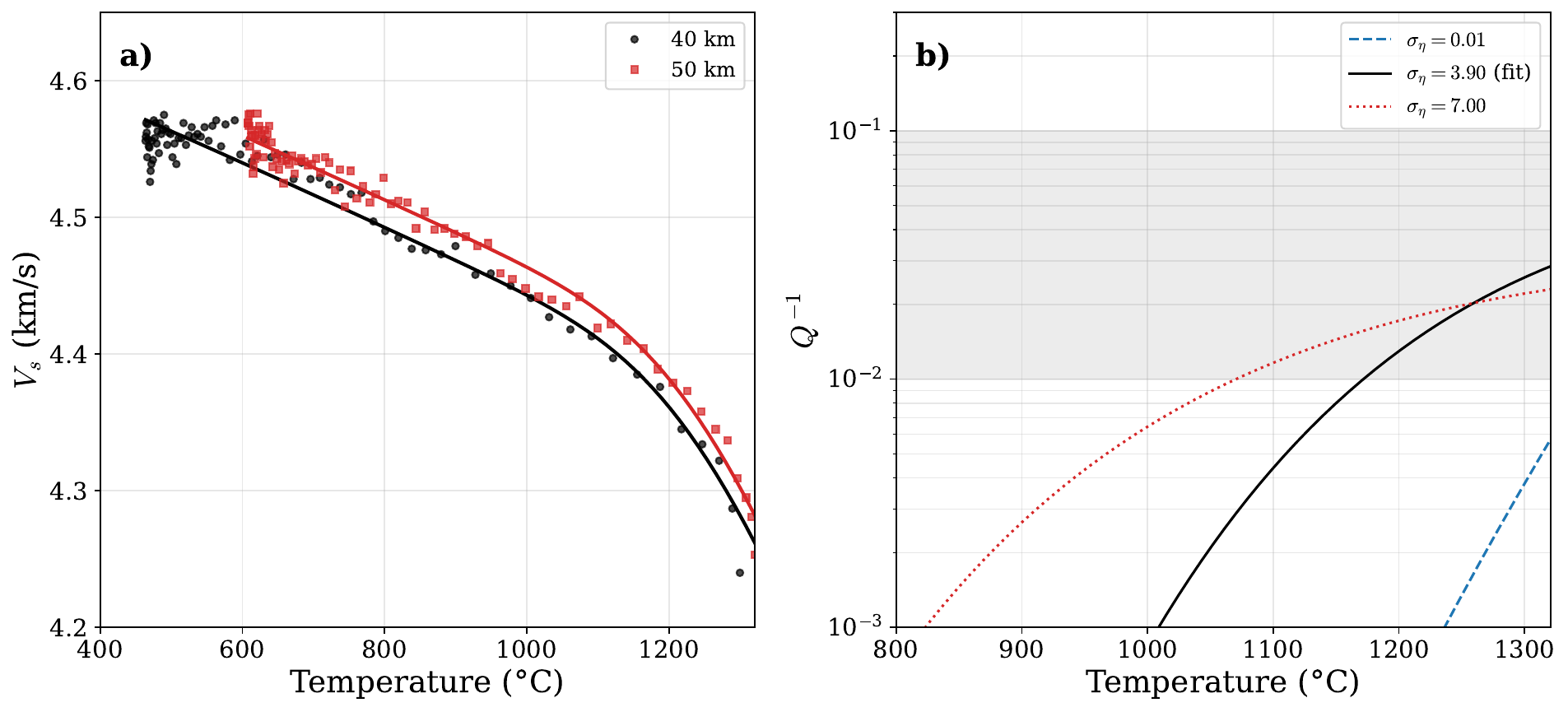}
    \caption{\textbf{Surface-wave velocity and attenuation inferred by the reduced-order distributed-viscosity EAGBS model.}
    \textbf{(a)} Best-fitting velocity--temperature relationship (solid curves) compared with the seismic data compiled by \citeA{priestleyThermalAnisotropicStructure2024} at 40 and 50 km depth (symbols). 
    \textbf{(b)} Corresponding model prediction for attenuation, shown as inverse quality factor $Q^{-1}$, as a function of temperature. The best-fitting case ($\sigma_\eta=3.90$) is compared with fixed narrow ($\sigma_\eta=0.01$) and broad ($\sigma_\eta=7.00$) distribution cases obtained by varying $\sigma_\eta$ while holding all other parameters fixed at their best-fit values. The shaded band marks the order-of-magnitude range $Q^{-1}$$\sim$$10^{-2}$ inferred for the asthenosphere. Narrow viscosity distributions produce insufficient attenuation, whereas broader distributions yield seismically relevant values.}
    \label{fig:fitting}
\end{figure}

The resulting fit and the corresponding attenuation calculated at a 40~s period are shown in Figure~\ref{fig:fitting}. The fit shows that the distributed-viscosity EAGBS model can account for the nonlinear high-temperature reduction in $V_s$, while also generating attenuation of the order required by asthenospheric observations. In particular, the narrow-distribution case produces insufficient attenuation, whereas the fitted broader distribution, with $\sigma_\eta=3.90$, yields attenuation of order $Q^{-1}\sim10^{-2}$ over asthenospheric temperatures. Repeating the same plausibility fit for assumed periods from 40 to 100~s leaves the velocity misfit effectively unchanged and changes the fitted viscosity distribution width only weakly, from $\sigma_\eta=3.90$ to 3.65; the sensitivity in period is summarized in \ref{app:fitting_model} and Table~\ref{tab:period_sensitivity}.

The best-fitting activation energy is $E_a=429$~kJ/mol, compared with the 666~kJ/mol reported by \citeA{quOnsetAnelasticBehavior2024}. The fitted viscosity-distribution width, $\sigma_\eta=3.90$, indicates substantial grain-boundary rheological heterogeneity in a model that reproduces the velocity reduction and generates seismically relevant attenuation. This is consistent with the hypothesis developed in Sections~4 and 5.2 that dry olivine may sample grain-boundary viscosities spanning several orders of magnitude.

The seismic comparison is a plausibility test rather than a unique inversion for grain-boundary properties. The fitted elastic coefficients are of the same order of magnitude as laboratory values and are used here as effective parameters, not as independent measurements of olivine elasticity (\ref{app:fitting_model}). Nevertheless, the results show that distributed-viscosity EAGBS can produce both the nonlinear high-temperature velocity reduction and attenuation of order $Q^{-1}\sim10^{-2}$, supporting its potential contribution to upper-mantle seismic anelasticity.

Other candidate mechanisms have also been proposed to explain the missing dispersion and attenuation in the upper asthenosphere. One interpretation invokes poroelastic effects \cite{mavkoVelocityAttenuationPartially1980,takeiEffectPoreGeometry2002} and enhanced anelasticity \cite{faulShearWaveAttenuation2004} associated with partial melting. However, geochemical constraints suggest that the melt fraction in the upper mantle is very small \cite{mckenzieConstraintsMeltGeneration2000}, making it difficult to attribute the observed large dispersion to melt alone \cite{takeiPhaseFieldModelingGrain2019}. Another possible explanation is grain-boundary premelting \cite{yamauchiApplicationPremeltingModel2020a}, motivated by the observation of a high-frequency attenuation peak in borneol rock-analogue experiments \cite{takeiTemperatureGrainSize2014, yamauchiPolycrystalAnelasticityNearsolidus2016}. This peak exhibits scaling relationships different from those of the EAGBS peak considered in this study. However, the physical mechanism by which premelting enhances anelastic behaviour and produces such a peak remains unclear. Further theoretical and experimental work is therefore needed to clarify the role of premelt in the upper asthenosphere.

\subsection{Scope and Limitations of the Model}\label{sec:limitations}

Several limitations bound the interpretation of the results presented above.

The first concerns the assumed interfacial rheology. Following the EAGBS treatments of \citeA{ghahremaniEffectGrainBoundary1980} and \citeA{leeAnelasticityGrainBoundary2010}, the present model represents tangential sliding by a linear Newtonian interface. Its applicability to real olivine grain boundaries remains uncertain, and neither the numerical results nor the velocity fit of Section~5.3 constitutes a validation of this constitutive choice: the fit tests whether a distributed-viscosity EAGBS response is capable of reproducing the observations, not whether the underlying interfacial law is correct. Testing nonlinear or otherwise more general interfacial laws is a natural extension of this framework, and the distributed-viscosity treatment developed here would apply equally to a distribution over the parameters of such a law.

The second concerns the frequency range over which the model is valid. Diffusionally assisted grain-boundary sliding is not included. On the estimate of \citeA{ashbyBoundaryDefectsAtomistic1972}, the ratio $\tau_e/\tau_M$ is of order $10^{-8}$ for $1~\mu\mathrm{m}$ grains and $10^{-14}$ for $1~\mathrm{mm}$ grains \cite{jacksonGrainsizesensitiveSeismicWave2002,takeiTemperatureGrainSize2014,rudgeViscoelasticRheologyTransient2026}, placing the EAGBS relaxation time many orders of magnitude below the Maxwell time. This separation justifies treating the EAGBS contribution over its own frequency interval while leaving the lower-frequency transition outside the model. It does not preclude a broad diffusionally assisted attenuation background between $\tau_e$ and $\tau_M$ \cite{leeStressConcentrationsDiffusionally2011,morrisDiffusionallyAssistedGrainboundary2009}. A combined treatment would be appropriate for a full mantle rheology while the present calculation tests the contribution that distributed-viscosity EAGBS can make on its own. The auxiliary laboratory comparison in \ref{app:dry_joint_fit} illustrates how the two contributions can be combined.

The third concerns the microstructural representation. The calculations are restricted to 2-D isotropic elastic aggregates with prescribed statistical distributions of grain-boundary viscosity, and do not include crystallographic elastic anisotropy or spatial correlation between the properties of neighbouring boundaries. Extending the framework to 3-D microstructures, and constraining the physically realistic distribution of grain-boundary viscosities in olivine, are the key next steps.

\section{Conclusions}

This study tests whether microstructural heterogeneity can reconcile the classical EAGBS prediction of a localized Debye-like attenuation peak with the weak or poorly resolved peak observed in dry olivine experiments. The numerical results show that different forms of heterogeneity have different effects on the macroscopic spectrum.

Irregular grain geometry substantially changes the baseline EAGBS response relative to the regular hexagonal benchmark, but increasing grain-size variance within irregular Voronoi aggregates produces only modest changes in relaxed modulus and peak height, with little spectral broadening. Within the present model framework, grain-size heterogeneity alone is therefore insufficient to account for the broad, weak attenuation background observed in dry olivine. By contrast, a broad distribution of grain-boundary viscosities progressively suppresses and broadens the Debye-like peak into a weak background extending across a wide frequency interval. The bin-resolved decomposition shows that this broadening does not require intrinsically non-linear local behaviour. Instead, it arises from the superposition of many localized relaxation processes with different characteristic timescales. The approximately additive scaling of integrated dissipation with the boundary length in each bin then motivates a reduced-order 0-D description of the aggregate response.

These results imply that the absence of a pronounced EAGBS peak in dry olivine does not necessarily indicate the absence of EAGBS itself. If dry olivine samples a sufficiently broad distribution of grain-boundary viscosities, the EAGBS contribution may be distributed over such a wide range of relaxation times that it appears experimentally only as part of a broad attenuation background. Conversely, if hydration or other chemical modification narrows the effective grain-boundary viscosity distribution, a more distinct peak may emerge. The same mechanism may also be relevant for upper-mantle seismology: even when viscosity heterogeneity suppresses the amplitude of the loss peak, the broadened EAGBS spectrum can still generate attenuation of order \(Q^{-1}\sim10^{-2}\) and associated velocity reductions of order a few percent relative to the low-temperature elastic trend. EAGBS therefore remains a plausible candidate mechanism for upper-mantle seismic attenuation and dispersion, provided that grain-boundary viscosities are sufficiently heterogeneous.

The limitations of this approach are set out in Section~\ref{sec:limitations}. In particular, the seismic comparison should be interpreted as a plausibility test within the EAGBS regime rather than as a unique inversion for mantle grain-boundary properties, or as a complete rheology spanning the transition to diffusionally accommodated sliding. Nevertheless, the results presented here identify grain-boundary viscosity heterogeneity as a potentially fundamental control on the macroscopic expression of EAGBS in polycrystalline mantle rocks.

\appendix

\section{Governing equations} \label{App:equations}

Grain interiors are modeled as isotropic, linearly elastic solids in the quasistatic limit. Neglecting inertial terms, the local balance of linear momentum is
\begin{equation}
    \Div \tensor{\sigma} = \mathbf{0}
    \qquad \text{in each grain}, 
    \label{eq:momentum}
\end{equation}
where $\tensor{\sigma}$ is the Cauchy stress tensor. The constitutive relation is
\begin{equation}
    \sigma_{ij} = \lambda \varepsilon_{kk} \delta_{ij} + 2\mu \varepsilon_{ij},
    \label{eq:hooke}
\end{equation}
with strain tensor
\begin{equation}
    \varepsilon_{ij} = \frac{1}{2}\left(\pd{u_i}{x_j} + \pd{u_j}{x_i}\right),
    \label{eq:strain}
\end{equation}
where $\vect{u}$ is the displacement field, and $\lambda$ and $\mu$ are the Lam\'e parameters.

Grain boundaries are represented as infinitesimally thin viscous interfaces. Tangential sliding is governed by
\begin{equation}
    \delta \sigma_{ns} = \eta \left[\pd{u_s}{t}\right],
    \label{eq:gb_viscous}
\end{equation}
while continuity of normal displacement is enforced by
\begin{equation}
    [u_n] = 0.
    \label{eq:gb_normal}
\end{equation}

Here $\sigma_{ns}$ is the shear traction resolved on the grain boundary, $\delta$ is the thickness of the viscous intergranular layer, $\eta$ is its viscosity, and $[\cdot]$ denotes the jump across the boundary. The subscripts $n$ and $s$ denote the components normal and tangential to the grain boundary, respectively. Equation~\eqref{eq:gb_viscous} corresponds to tangential viscous sliding in a thin intergranular film \cite{rajGrainBoundarySliding1971}, whereas equation~\eqref{eq:gb_normal} prevents boundary opening.

\section{Non-dimensionalization}\label{App:Non-Dim}

To identify the relevant control parameters, the equations are non-dimensionalized using a characteristic grain size $a$, a characteristic displacement amplitude $u_0$, and a reference grain-boundary viscosity $\eta_0$. We define
\begin{subequations}
\begin{align}
    \vect{u} &= u_0 \vect{u}^{\prime}, \\
    \tensor{\sigma} &= \frac{\mu u_0}{a}\tensor{\sigma}^{\prime}, \\
    t &= t_{\eta} t^{\prime}, \\
    \tensor{\varepsilon} &= \frac{u_0}{a}\tensor{\varepsilon}^{\prime},
\end{align}
\label{eq:nondim}
\end{subequations}
where the characteristic sliding time scale is
\begin{equation}
    t_{\eta} = \frac{\eta_0 a}{\delta\mu}.
    \label{eq:sliding_time_app}
\end{equation}
After substitution of equations~\eqref{eq:nondim}--\eqref{eq:sliding_time_app} and dropping primes for clarity, the governing equations become
\begin{equation}
    \Div \tensor{\sigma} = \mathbf{0}
    \qquad \text{in each grain},
    \label{eq:nondim_momentum}
\end{equation}
together with the grain-boundary conditions
\begin{subequations}
\begin{align}
    \sigma_{ns} &= \frac{\eta}{\eta_0}\left[\pd{u_s}{t}\right],
    \label{eq:nondim_slide}\\
    [u_n] &= 0.
    \label{eq:nondim_normal}
\end{align}
\label{eq:nondim_gb}
\end{subequations}

Under time-harmonic loading of the form $\mathrm{e}^{i\omega t}$, the time derivative is replaced by multiplication by $i\omega$, giving
\begin{equation}
    \sigma_{ns} = i\omega \frac{\eta}{\eta_0}[u_s].
    \label{eq:harmonic_slide}
\end{equation}
Here $\omega$ is the non-dimensional angular frequency.

\section{Discretization}\label{App:Discretization}

The governing equations are solved on a representative volume element (RVE) containing multiple grains separated by viscous interfaces. Because the displacement field is discontinuous across grain boundaries, the RVE is partitioned into subdomains, each containing a single grain or grain fragment, and the interior elasticity problem is discretized in $H^1$ on each subdomain. Here, $H^1$ denotes the Sobolev space of functions whose values and first weak derivatives are square-integrable, which is the natural finite-element space for continuous displacement fields within each elastic grain. The grain-interior displacement is discretized with quadratic ($P^2$) elements, while the grain-boundary tractions are discretized with linear ($P^1$) elements.

\subsection{Weak form in grain interiors}

For a grain occupying domain $\Omega_i$, we test equation~\eqref{eq:nondim_momentum} against a vector test function $\vect{v}\in H^1(\Omega_i)$. Integration by parts yields
\begin{equation}
    \int_{\Omega_i} \tensor{\sigma}^{*}(\vect{u}) : \tensor{\varepsilon}(\vect{v})\, \mathrm{d}V
    - \int_{\partial \Omega_i} \vect{v}\cdot \tensor{\sigma}^{*}(\vect{u})\cdot \vect{n}_i\, \mathrm{d}S
    = 0,
    \label{eq:weak_grain}
\end{equation}
where $(\cdot)^*$ denotes complex conjugation, consistent with the complex harmonic representation.

\subsection{Weak form on grain boundaries}
\label{sec:weak_gb}

Consider two neighboring grains $\Omega_i$ and $\Omega_j$ sharing an interface
\[
\Gamma_{\mathrm{GB}} = \partial \Omega_i \cap \partial \Omega_j.
\]
Combining the boundary terms from the two grains gives
\begin{equation}
    \int_{\partial\Omega_i} \vect{v}\cdot\tensor{\sigma}^{*}(\vect{u})\cdot\vect{n}_i\, \mathrm{d}S
    + \int_{\partial\Omega_j} \vect{v}\cdot\tensor{\sigma}^{*}(\vect{u})\cdot\vect{n}_j\, \mathrm{d}S
    =
    \int_{\Gamma_{\mathrm{GB}}} [\vect{v}] \cdot \tensor{\sigma}^{*}(\vect{u})\cdot \vect{n}_j\, \mathrm{d}S,
\end{equation}
where $\vect{n}_j=-\vect{n}_i$ and $[\vect{v}]=\vect{v}_i-\vect{v}_j$.

Let $\vect{n}$ and $\vect{s}$ denote the local unit normal and tangential directions on the interface. The traction vector $\tensor{\sigma}\cdot\vect{n}$ is decomposed into normal and tangential components,
\begin{equation}
    t_n = \vect{n}\cdot \tensor{\sigma}\cdot \vect{n},
    \qquad
    t_s = \vect{s}\cdot \tensor{\sigma}\cdot \vect{n}.
\end{equation}
The interface contribution may then be written as
\begin{equation}
    \int_{\Gamma_{\mathrm{GB}}} [\vect{v}] \cdot \tensor{\sigma}^{*}(\vect{u})\cdot \vect{n}\, \mathrm{d}S
    =
    \int_{\Gamma_{\mathrm{GB}}}
    \left(
    [v_s]\, t_s^{*}
    +
    [v_n]\, t_n^{*}
    \right)\mathrm{d}S.
\end{equation}

The grain-boundary conditions are imposed weakly by treating $t_s$ and $t_n$ as additional unknown interface tractions, with corresponding test functions $r_s$ and $r_n$. From equations~\eqref{eq:harmonic_slide} and \eqref{eq:nondim_normal}, we obtain
\begin{equation}
    \frac{i}{\omega}\frac{\eta_0}{\eta}
    \int_{\Gamma_{\mathrm{GB}}} t_s^{*} r_s \,\mathrm{d}S
    =
    \int_{\Gamma_{\mathrm{GB}}} [u_s]^{*} r_s \,\mathrm{d}S,
\end{equation}
and
\begin{equation}
    \int_{\Gamma_{\mathrm{GB}}} [u_n]^{*} r_n \,\mathrm{d}S = 0.
\end{equation}

Collecting the interior and interface contributions, the weak form becomes
\begin{equation}
\begin{aligned}
&\sum_i \int_{\Omega_i} \tensor{\sigma}^{*}(\vect{u}) : \tensor{\varepsilon}(\vect{v})\, \mathrm{d}V \\
&\quad
- \int_{\Gamma_{\mathrm{GB}}} [v_s]\, t_s^{*}\, \mathrm{d}S
- \int_{\Gamma_{\mathrm{GB}}} [u_s]^{*} r_s\, \mathrm{d}S
+ \frac{i}{\omega}\frac{\eta_0}{\eta}
  \int_{\Gamma_{\mathrm{GB}}} t_s^{*} r_s\, \mathrm{d}S \\
&\quad
- \int_{\Gamma_{\mathrm{GB}}} [v_n]\, t_n^{*}\, \mathrm{d}S
- \int_{\Gamma_{\mathrm{GB}}} [u_n]^{*} r_n\, \mathrm{d}S \\
&\quad
+ \text{(RVE boundary terms)}
= 0.
\end{aligned}
\label{eq:weak_total_preperiodic}
\end{equation}

\subsection{Periodic boundary conditions on the RVE}

For a periodic RVE, the total displacement is decomposed into a macroscopic affine part and a periodic fluctuation,
\begin{equation}
    \vect{u}_{\mathrm{total}} = \vect{U} + \vect{u},
\end{equation}
where $\vect{U}$ represents the imposed macroscopic deformation and $\vect{u}$ is periodic over the RVE.

Following the micropolar homogenization framework of \citeA{rudgeMicropolarContinuumModel2021,rudgeViscoelasticRheologyTransient2026}, the displacement jump between neighboring RVEs may be written as
\begin{equation}
    [\vect{U}] = \tensor{\Gamma}\cdot\vect{R} + (\tensor{K}\cdot\vect{R})\times\vect{d},
\end{equation}
where $\vect{R}$ joins the centroids of neighboring RVEs and $\vect{d}$ joins the centroid of the neighboring RVE to a point on the boundary. In the present study, we adopt a Cauchy-continuum upscaling and neglect granular-scale rotational degrees of freedom, so that $\tensor{K}=\tensor{0}$ and $\tensor{\Gamma}$ is symmetric. The imposed jump therefore reduces to
\begin{equation}
    [\vect{U}] = \tensor{\Gamma}\cdot\vect{R}.
\end{equation}

Applying the same interface treatment as in Section~\ref{sec:weak_gb} to the periodic RVE boundary gives
\begin{equation}
\begin{aligned}
&\sum_i \int_{\Omega_i} \tensor{\sigma}^{*}(\vect{u}) : \tensor{\varepsilon}(\vect{v})\, \mathrm{d}V \\
&\quad
- \int_{\Gamma_{\mathrm{GB}}}
\left(
[v_s]\, t_s^{*}
+
[v_n]\, t_n^{*}
\right)\mathrm{d}S
- \int_{\Gamma_{\mathrm{GB}}}
\left(
[u_s]^{*} r_s
+
[u_n]^{*} r_n
\right)\mathrm{d}S \\
&\quad
+ \frac{i}{\omega}\frac{\eta_0}{\eta}
\int_{\Gamma_{\mathrm{GB}}} t_s^{*} r_s\, \mathrm{d}S \\
&\quad
- \int_{\Gamma_{\mathrm{top}}\cup\Gamma_{\mathrm{right}}}
\left(
[v_s]\, t_s^{*}
+
[v_n]\, t_n^{*}
\right)\mathrm{d}S \\
&\quad
- \int_{\Gamma_{\mathrm{top}}\cup\Gamma_{\mathrm{right}}}
\left(
[u_s]^{*} r_s
+
[u_n]^{*} r_n
\right)\mathrm{d}S \\
&=
- \int_{\Gamma_{\mathrm{top}}\cup\Gamma_{\mathrm{right}}}
\left(\tensor{\Gamma}^{*}\cdot\vect{R}\cdot\vect{n}\right) r_n\, \mathrm{d}S
- \int_{\Gamma_{\mathrm{top}}\cup\Gamma_{\mathrm{right}}}
\left(\tensor{\Gamma}^{*}\cdot\vect{R}\cdot\vect{s}\right) r_s\, \mathrm{d}S.
\end{aligned}
\label{eq:weak_periodic}
\end{equation}

\section{Macroscopic energy measures and extraction of $G'(\omega)$, $G''(\omega)$, and binned contributions} \label{App:Homogenisation}

For each prescribed harmonic macroscopic loading and angular frequency $\omega$, the finite-element solution provides the complex displacement field and the corresponding grain-boundary tractions. The effective complex modulus is extracted from cycle-averaged energy measures.

Under harmonic loading, the displacement field is written as
\begin{equation}
    \vect{u}(\vect{x},t)
    =
    \Re\!\left[
        \hat{\vect{u}}(\vect{x})\,\mathrm{e}^{i\omega t}
    \right],
\end{equation}
where $\hat{\vect{u}} = \vect{u}^{\mathrm{R}} + i \vect{u}^{\mathrm{I}}$ is the complex displacement amplitude. The corresponding strain and stress amplitudes are decomposed as
\begin{equation}
    \hat{\tensor{\varepsilon}}
    =
    \tensor{\varepsilon}^{\mathrm{R}}
    +
    i\tensor{\varepsilon}^{\mathrm{I}},
    \qquad
    \hat{\tensor{\sigma}}
    =
    \tensor{\sigma}^{\mathrm{R}}
    +
    i\tensor{\sigma}^{\mathrm{I}}.
\end{equation}

The cycle-averaged elastic energy density stored in the grain interiors is
\begin{equation}
    w_{\mathrm{store}}
    =
    \frac{1}{2}
    \left(
        \tensor{\sigma}^{\mathrm{R}}:\tensor{\varepsilon}^{\mathrm{R}}
        +
        \tensor{\sigma}^{\mathrm{I}}:\tensor{\varepsilon}^{\mathrm{I}}
    \right).
\end{equation}
Integrating over the representative volume element (RVE) and normalizing by the RVE area $A$ gives the macroscopic stored elastic energy
\begin{equation}
    E_{\mathrm{store}}(\omega)
    =
    \frac{1}{2A}
    \sum_i
    \int_{\Omega_i}
    \left(
        \tensor{\sigma}^{\mathrm{R}}:\tensor{\varepsilon}^{\mathrm{R}}
        +
        \tensor{\sigma}^{\mathrm{I}}:\tensor{\varepsilon}^{\mathrm{I}}
    \right)\,\mathrm{d}V,
    \label{eq:E_store}
\end{equation}
where the sum is taken over all grains $\Omega_i$.

Dissipation occurs only on grain boundaries through viscous sliding. Let $\hat{t}_s = t_s^{\mathrm{R}} + i t_s^{\mathrm{I}}$ denote the complex tangential traction amplitude on a grain boundary $\Gamma^{(ij)}$. Using the viscous sliding law in equation~\eqref{eq:harmonic_slide}, the cycle-averaged energy dissipated on that boundary is
\begin{equation}
    E_{\mathrm{diss}}^{(ij)}(\omega)
    =
    \frac{1}{2A\,\omega}
    \frac{\eta_0}{\eta^{(ij)}}
    \int_{\Gamma^{(ij)}}
    \left|
        \hat{t}_s
    \right|^2
    \,\mathrm{d}S,
    \label{eq:E_diss_boundary}
\end{equation}
where
\begin{equation}
    \left|
        \hat{t}_s
    \right|^2
    =
    \left(t_s^{\mathrm{R}}\right)^2
    +
    \left(t_s^{\mathrm{I}}\right)^2.
\end{equation}
Summing over all internal grain boundaries gives the total macroscopic dissipation
\begin{equation}
    E_{\mathrm{diss}}(\omega)
    =
    \sum_{(i,j)} E_{\mathrm{diss}}^{(ij)}(\omega).
    \label{eq:E_diss_total}
\end{equation}

The effective response is obtained component-wise from the strain-energy measures. 
For a prescribed macroscopic strain component of amplitude $\Gamma_0$, the corresponding component of the effective complex stiffness tensor is
\begin{equation}
    C'_{ijkl}(\omega)
    =
    \frac{2E_{\mathrm{store}}(\omega)}{\Gamma_0^2},
    \qquad
    C''_{ijkl}(\omega)
    =
    \frac{2E_{\mathrm{diss}}(\omega)}{\Gamma_0^2},
    \label{eq:Cprime_Cdoubleprime}
\end{equation}
where the indices $ijkl$ denote the stiffness component associated with the imposed macroscopic strain mode. For the simple-shear loading used to calculate the shear response, this component is identified with the effective complex shear modulus,
\begin{equation}
    G^*(\omega)=G'(\omega)-iG''(\omega).
\end{equation}

To quantify how different viscosity classes contribute to the macroscopic attenuation spectrum, the set of grain boundaries is partitioned into bins according to their assigned viscosity $\eta^{(ij)}$. Let $\mathcal{B}_k$ denote the set of boundaries in viscosity bin $k$. The dissipative contribution of that bin is then
\begin{equation}
    E_{\mathrm{diss}}^{(k)}(\omega)
    =
    \sum_{(i,j)\in\mathcal{B}_k}
    E_{\mathrm{diss}}^{(ij)}(\omega),
    \label{eq:E_diss_bin}
\end{equation}
and the corresponding contribution to the imaginary modulus is
\begin{equation}
    C''^{(k)}(\omega)
    =
    \frac{2E_{\mathrm{diss}}^{(k)}(\omega)}{\Gamma_0^2}.
    \label{eq:C_imag_bin}
\end{equation}
By construction,
\begin{equation}
    C''(\omega)
    =
    \sum_k C''^{(k)}(\omega),
\end{equation}
so the total attenuation spectrum is decomposed into additive contributions from different grain-boundary viscosity classes.

This decomposition is used in Section~4 to show that broad macroscopic attenuation arises from the superposition of many localized dissipation spectra associated with boundaries relaxing at different characteristic time scales.

\section{Reduced-order model used to fit seismic velocity--temperature data}
\label{app:fitting_model}

To examine whether distributed-viscosity EAGBS can reproduce the nonlinear high-temperature reduction in seismic velocity, we fitted the reduced-order model described in Section~\ref{sec:0-D} to the surface-wave shear-velocity data used by \citeA{priestleyThermalAnisotropicStructure2024}. The fit was performed for the 40 and 50 km depth datasets.

\subsection{Low-temperature elastic parameterization}
Following \citeA{priestleyThermalAnisotropicStructure2024}, the unrelaxed shear modulus was taken to vary linearly with temperature and pressure,
\begin{equation}
    \mu(T,P) = \mu_0 + \frac{\partial \mu}{\partial T} T + \frac{\partial \mu}{\partial P} P,
    \label{eq:mu_fit}
\end{equation}
where $\mu_0$, $\partial \mu/\partial T$, and $\partial \mu/\partial P$ are fitting parameters. Here $T$ is in $^\circ$C and $P$ is in GPa.

Density was prescribed as
\begin{equation}
    \rho(T,P) = \rho_0 \left(1 - \alpha T + \frac{P}{K_T}\right),
    \label{eq:rho_fit}
\end{equation}
with fixed constants $\rho_0 = 3213~\mathrm{kg\,m^{-3}}$, $\alpha = 4.07\times10^{-5}~^\circ\mathrm{C}^{-1}$, and $K_T = 115$ GPa \cite{isaakHightemperatureElasticityIronbearing1992}.

Rather than prescribing pressure from a fixed depth gradient, pressure was computed self-consistently from hydrostatic balance,
\begin{equation}
    \frac{\mathrm{d}P}{\mathrm{d}z} = \rho(T,P)\,g,
    \label{eq:hydrostatic_fit}
\end{equation}
where $g = 9.81~\mathrm{m\,s^{-2}}$. The model is evaluated pointwise at each $(T,z)$ pair: the temperature $T$ is treated as uniform over the column $[0,z]$ when integrating \eqref{eq:hydrostatic_fit}, and the resulting $P(z,T)$ is used to evaluate $\mu(T,P)$ and $\tau(T,P)$. This isothermal-column approximation introduces negligible error (about $5\%$ error in local density and about $2\%$ error in the integrated pressure) because $\alpha T \ll 1$ over the mantle temperature range. Substituting equation~\eqref{eq:rho_fit} into \eqref{eq:hydrostatic_fit} gives the analytic solution
\begin{equation}
    P(z,T)
    =
    \frac{a}{b}\left(\mathrm{e}^{bz}-1\right),
    \label{eq:self_consistent_pressure}
\end{equation}
where
\begin{equation}
    a = \rho_0 (1-\alpha T) g,
    \qquad
    b = \frac{\rho_0 g}{K_T},
\end{equation}
with $P$ in Pa, $z$ in m, and $K_T$ expressed in Pa in equation~\eqref{eq:self_consistent_pressure}. In the implementation, pressure is converted to GPa before evaluating equation~\eqref{eq:mu_fit}.

The corresponding unrelaxed shear-wave velocity is
\begin{equation}
    V_{s,u}(T,P) = \sqrt{\frac{\mu(T,P)}{\rho(T,P)}}.
    \label{eq:Vs_unrelaxed_fit}
\end{equation}

\subsection{Temperature-dependent EAGBS timescale}

The EAGBS relaxation time was parameterized by scaling from a laboratory reference state:
\begin{equation}
    \tau(T,P)
    =
    \tau_{\mathrm{ref}}
    \left(\frac{d}{d_{\mathrm{ref}}}\right)
    \left(\frac{\mu_{\mathrm{ref}}}{\mu(T,P)}\right)
    \left(\frac{T_K}{T_{\mathrm{ref},K}}\right)
    \exp\!\left[
        \frac{E_a}{R}
        \left(
            \frac{1}{T_K} - \frac{1}{T_{\mathrm{ref},K}}
        \right)
    \right],
    \label{eq:tau_fit}
\end{equation}
where $T_K=T+273.15$ is temperature in Kelvin, $R$ is the gas constant, $E_a$ is the activation energy, and
\begin{equation}
    \mu_{\mathrm{ref}} = \mu(T_{\mathrm{ref}},P_{\mathrm{ref}}).
\end{equation}
The factor $T_K/T_{\mathrm{ref},K}$ follows from the assumed atomistic form $\eta_{\mathrm{gb}}\propto T_K/D_{\mathrm{gb}}$ for grain-boundary sliding viscosity \cite{ashbyBoundaryDefectsAtomistic1972,quOnsetAnelasticBehavior2024}. With $D_{\mathrm{gb}}\propto\exp[-E_a/(RT_K)]$ and $\tau_e\propto\eta_{\mathrm{gb}}d/\mu$ (equation~\ref{eq:sliding_time}), this form gives the temperature, grain-size, and modulus factors in equation~\ref{eq:tau_fit}. 
In the fit, the fixed reference values were
\begin{equation}
    d = 10^{-3}\ \mathrm{m},\qquad
    d_{\mathrm{ref}} = 5\times10^{-6}\ \mathrm{m},\qquad
    T_{\mathrm{ref}} = 900^\circ\mathrm{C},\qquad
    P_{\mathrm{ref}} = 0.2\ \mathrm{GPa}.
\end{equation}
The parameter $\tau_{\mathrm{ref}}$ is therefore an effective reference sliding time at the laboratory anchor state, and was fitted in logarithmic form through $\log_{10}\tau_{\mathrm{ref}}$.

\subsection{Distributed-viscosity standard-linear-solid representation} \label{App: E3}

The dimensionless complex modulus reduction factor was represented as a parallel superposition of Debye elements,
\begin{equation}
    M^*(\omega;T,P)
    =
    1
    -
    \Delta
    \int_{-\infty}^{\infty}
    \frac{p(\ln \hat{\tau})}{1+i\omega \tau(T,P)\hat{\tau}}
    \,\mathrm{d}(\ln \hat{\tau}),
    \label{eq:Mstar_continuous_fit}
\end{equation}
where $\Delta$ is the total relaxation strength and $p(\ln\hat{\tau})$ is a normal distribution in $\ln \hat{\tau}$ with standard deviation $\sigma_\eta$,
\begin{equation}
    p(\ln \hat{\tau})
    =
    \frac{1}{\sigma_\eta\sqrt{2\pi}}
    \exp\!\left[
        -\frac{(\ln \hat{\tau})^2}{2\sigma_\eta^2}
    \right].
    \label{eq:lognormal_tau_pdf}
\end{equation}
In practice, this integral was evaluated numerically as a weighted quadrature over $\ln \hat{\tau}$, truncated to $\pm 6\sigma_\eta$ and discretized using 2001 points. The relaxation strength was fixed at
\begin{equation}
    \Delta = 0.28,
\end{equation}
motivated by the finite-element results for irregular Voronoi aggregates.

For a seismic period $T_p$, with angular frequency $\omega=2\pi/T_p$, the dispersed shear-wave velocity was computed as
\begin{equation}
    V_s(T,P;\omega)
    =
    V_{s,u}(T,P)\,
    \Re\!\left[\sqrt{M^*(\omega;T,P)}\right].
    \label{eq:Vs_dispersed_fit}
\end{equation}

The inverse quality factor was evaluated from the complex compliance,
\begin{equation}
    J^*(\omega;T,P)=\frac{1}{M^*(\omega;T,P)},
\end{equation}
through
\begin{equation}
    Q^{-1}(\omega;T,P)
    =
    -\frac{\Im(J^*)}{\Re(J^*)}.
    \label{eq:Qinv_fit}
\end{equation}

\subsection{Fitting procedure}

The model was fitted by nonlinear least squares to the observed shear-wave velocities. The residual vector was defined by
\begin{equation}
    r_i = V_s^{\mathrm{model}}(T_i,z_i)-V_{s,i}^{\mathrm{obs}}.
\end{equation}

The inversion was performed in two stages. In Stage 1, only the low-temperature data ($T\le 900^\circ$C) at 40 and 50 km were used to estimate the low-temperature elastic parameters $\mu_0$, $\partial\mu/\partial T$, and $\partial\mu/\partial P$, while holding the viscoelastic parameters fixed at their initial values. In Stage 2, all six parameters were fitted simultaneously using the 40 and 50 km datasets. The seismic period used in the velocity fit was chosen at $T_p=40$ s.

The fitted parameter vector is therefore
\begin{equation}
    \mathbf{x}
    =
    \left(
    \mu_0,\,
    \frac{\partial\mu}{\partial T},\,
    \frac{\partial\mu}{\partial P},\,
    \log_{10}\tau_{\mathrm{ref}},\,
    E_a,\,
    \sigma_\eta
    \right).
\end{equation}

Note that $\mu_0$ is the extrapolated shear modulus at $0^\circ\mathrm{C}$, so small changes in $\partial\mu/\partial T$ can produce large changes in the fitted intercept. Similarly, strong trade-offs exist among $E_a$, $\tau_{\mathrm{ref}}$, and $\sigma_\eta$.

For comparison, laboratory reference values are $\mu_0=78.7$~GPa, $\partial\mu/\partial T=-0.0136$~GPa~K$^{-1}$, and $\partial\mu/\partial P=1.8$ \cite{isaakHightemperatureElasticityIronbearing1992,priestleyThermalAnisotropicStructure2024}. The fitted values in Table~\ref{tab:fitting_parameters} have comparable magnitudes, but they are effective parameters of the seismic plausibility fit rather than independent laboratory determinations of olivine elasticity.

\begin{table}
    \centering
    \begin{tabular}{lll}
        \hline
        Parameter & Symbol & Best-fit value \\
        \hline
        Reference shear modulus intercept & $\mu_0$ & $67.9120$ GPa \\
        Temperature derivative of shear modulus & $\partial\mu/\partial T$ & $-9.30735\times10^{-3}$ GPa $^\circ$C$^{-1}$ \\
        Pressure derivative of shear modulus & $\partial\mu/\partial P$ & $2.41758$ GPa GPa$^{-1}$ \\
        Logarithm of reference relaxation time & $\log_{10}\tau_{\mathrm{ref}}$ & $5.09999$ \\
        Reference relaxation time & $\tau_{\mathrm{ref}}$ & $1.26\times10^{5}$ s \\
        Activation energy & $E_a$ & $429.436$ kJ mol$^{-1}$ \\
        Width of log-normal timescale distribution & $\sigma_\eta$ & $3.89721$ \\
        Fixed relaxation strength & $\Delta$ & $0.28$ \\
        Fixed seismic period for velocity fit & $T_p$ & $40$ s \\
        Fixed mantle grain size & $d$ & $1.0\times10^{-3}$ m \\
        Fixed laboratory reference grain size & $d_{\mathrm{ref}}$ & $5.0\times10^{-6}$ m \\
        Fixed reference temperature & $T_{\mathrm{ref}}$ & $900^\circ$C \\
        Fixed reference pressure & $P_{\mathrm{ref}}$ & $0.2$ GPa \\
        \hline
        Overall root-mean-square error (RMSE) &  & $0.012505$ km s$^{-1}$ \\
        Overall mean absolute error (MAE) &  & $0.009489$ km s$^{-1}$ \\
        RMSE at 40 km &  & $0.014316$ km s$^{-1}$ \\
        RMSE at 50 km &  & $0.010355$ km s$^{-1}$ \\
        \hline
    \end{tabular}
    \caption{\textbf{Best-fitting parameters for the reduced-order distributed-viscosity EAGBS model.} The fit was performed to the 40 and 50 km shear-wave velocity datasets. }
    \label{tab:fitting_parameters}
\end{table}

\clearpage
\subsection{Sensitivity to the assumed seismic period}

To test whether the single period adopted in the reference fit controls the main inference, we repeated the same two-stage fitting procedure at 5~s intervals from 40 to 100~s. Representative results are summarized in Table~\ref{tab:period_sensitivity}. The elastic parameters and the velocity misfit are unchanged at the displayed precision. The fitted activation energy decreases by 5.8\%, from 429.44 to 404.41~kJ~mol$^{-1}$, while the fitted distribution width decreases by 6.4\%, from $\sigma_\eta=3.897$ to 3.646. Thus, over 40--100~s, the fitted viscosity-distribution width changes by only 6.4\%, indicating modest sensitivity to the assumed period.

\begin{table}[htbp]
    \centering
    \begin{tabular}{lccc}
        \hline
        Assumed period, $T_p$ (s) & 40 & 70 & 100 \\
        \hline
        Activation energy, $E_a$ (kJ mol$^{-1}$) & 429.44 & 414.15 & 404.41 \\
        Distribution width, $\sigma_\eta$ & 3.897 & 3.744 & 3.646 \\
        Velocity RMSE (km s$^{-1}$) & 0.012505 & 0.012505 & 0.012505 \\
        Velocity MAE (km s$^{-1}$) & 0.009489 & 0.009489 & 0.009490 \\
        \hline
    \end{tabular}
    \caption{\textbf{Sensitivity of the reduced-order EAGBS plausibility fit to the assumed seismic period.} The same two-stage fitting procedure was repeated from 40 to 100~s in 5~s increments; representative periods are shown.}
    \label{tab:period_sensitivity}
\end{table}

\section{Auxiliary joint fit to dry-olivine modulus and attenuation}\label{app:dry_joint_fit}

This exercise tests whether the modulus decrement predicted by our 2-D finite-element aggregates necessarily implies a visible EAGBS peak in dry-olivine measurements. Its fixed strength is illustrative, since the 2-D value has not been calibrated for 3-D olivine. We used the 59 paired modulus--attenuation measurements for each of the A1802 and A1928 specimens reported by \citeA{quOnsetAnelasticBehavior2024}. The same log-normal distributed-viscosity EAGBS modulus factor $M^*_{\mathrm{EAGBS}}$ as in equation~\eqref{eq:Mstar_continuous_fit} was combined with the generalized-Burgers DAGBS background at the level of normalized complex compliance,
\begin{equation}
    j^*_{\mathrm{total}}=j^*_{\mathrm{DAGBS}}+\left[\left(M^*_{\mathrm{EAGBS}}\right)^{-1}-1\right].
    \label{eq:joint_compliance_dry}
\end{equation}
The modulus and $Q^{-1}$ were computed from this same total compliance. We fixed the EAGBS modulus decrement at $\Delta_M=0.28$, profiled $\sigma_\eta$, and refitted the EAGBS centre together with the DAGBS strength, exponent, upper relaxation time, and activation energy. Both observables entered the objective, using the 3\% modulus and 0.05-dex attenuation fitting scales of \citeA{quOnsetAnelasticBehavior2024}. For A1928 the DAGBS exponent was restricted to $0.35\leq\alpha\leq0.37$ to avoid the implausible background curvature obtained from an unconstrained decomposition.

\begin{figure}[htbp]
    \centering
    \includegraphics[width=\textwidth]{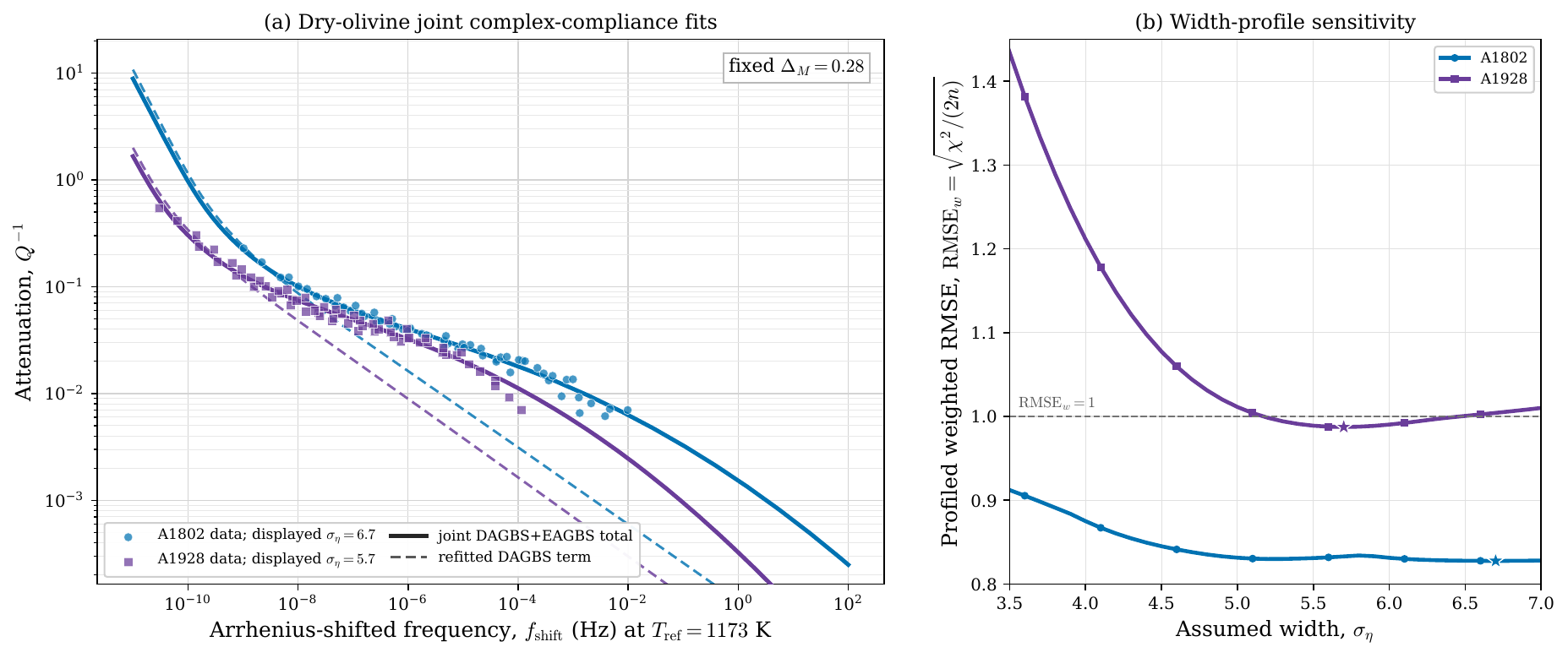}
    \caption{\textbf{Auxiliary joint DAGBS--EAGBS comparison with dry olivine.} \textbf{(a)} Measured attenuation for dry specimens A1802 and A1928 of \citeA{quOnsetAnelasticBehavior2024}, the total complex-compliance fits (solid curves), and the refitted DAGBS backgrounds (dashed curves). The EAGBS modulus decrement is fixed at the 2-D finite-element value $\Delta_M=0.28$; the displayed viscosity-distribution widths are $\sigma_\eta=6.7$ and 5.7, respectively.  \textbf{(b)} Weighted root-mean-square fitting error versus the assumed width. Stars mark the displayed profile minima.}
    \label{fig:dry_joint_fit}
\end{figure}

Figure~\ref{fig:dry_joint_fit} compares the fitted total attenuation with the data and the refitted DAGBS background. At the displayed profile minima, $\sigma_\eta=6.7$ for A1802 and 5.7 for A1928. The corresponding root-mean-square errors are 0.0159 and 0.0181 in normalized modulus and 0.0506 and 0.0586 dex in $\log_{10}Q^{-1}$; the weighted root-mean-square residuals are 0.828 and 0.987. No distinct EAGBS maximum is resolved in either fitted total spectrum. Although the shallow width profiles do not uniquely determine $\sigma_\eta$, the comparison demonstrates that the finite-element relaxation strength of 0.28 can coexist with the broad dry-olivine spectra without a prominent peak. The fitted widths are not intended to be the same as the seismic velocity-fit width of $\sigma_\eta=3.90$.

\section*{Data Availability Statement}

The finite-element homogenization and reduced-order modelling software, together with the documented workflow used to generate the geometries, numerical outputs, and figures reported in this study, is available in \cite{liZhengxuanLi471EAGBS_homogenisationV002026}. The archived release is distributed under the MIT license. All numerical results reported in this study can be reproduced using the archived software and workflow.

AI-assisted tools, including ChatGPT and Claude, were used to assist with language polishing in the manuscript, and with debugging, refactoring, and generating repetitive implementation code. All AI-assisted text and code were reviewed, edited, tested where applicable, and validated by the authors. The authors take full responsibility for all simulations, analyses, figures, interpretations, and conclusions presented in this manuscript.

\section*{Conflict of Interest disclosure}
The authors declare there are no conflicts of interest for this manuscript.

\acknowledgments
We thank Thomas Breithaupt, David Wallis, and Yasuko Takei for helpful comments and constructive discussions. We are grateful to Dan McKenzie for sharing the seismic velocity data used in \citeA{priestleyThermalAnisotropicStructure2024}. This work was performed using resources provided by the Cambridge Service for
Data Driven Discovery (CSD3), operated by the University of Cambridge Research
Computing Service (\url{https://www.csd3.cam.ac.uk}), provided by Dell EMC and
Intel using Tier-2 funding from the Engineering and Physical Sciences Research
Council (capital grant EP/T022159/1), and DiRAC funding from the Science and
Technology Facilities Council (\url{https://www.dirac.ac.uk}). No specific funding was received for this work.
\bibliography{references}

\end{document}